\documentclass[onecolumn, noshowpacs, amsmath,amssymb,nopreprintnumbers]{revtex4}
\usepackage{epsfig}
\usepackage[]{natbib}
\usepackage{appendix}
\usepackage{xspace}
\usepackage{color}
\usepackage{bbold}
\usepackage{tabu}

\usepackage{mathtools}
\usepackage{changes}     
\definechangesauthor[name={Author}, color=red]{Author}

\begin{document}

\title{
Fractional advection-diffusion-asymmetry equation}

\author{Wanli  Wang and Eli Barkai}
\affiliation{Department of Physics, Institute of Nanotechnology and Advanced Materials, Bar-Ilan University, Ramat-Gan
52900, Israel}

\begin{abstract}

Fractional kinetic equations employ non-integer calculus to model anomalous
relaxation and diffusion in many systems. While this approach is well explored,
it so far failed to describe an  important class of transport in disordered systems.  Motivated by work on contaminant
spreading in geological formations  we propose and investigate
a fractional advection-diffusion equation  describing the biased spreading packet.
 While usual
transport is described by diffusion and drift, we find a third term describing
symmetry breaking  which is omnipresent for transport in disordered systems.
Our work is based on continuous time random walks with a finite mean waiting
time and a diverging variance, a case that on the one hand is very common
and on the other was missing in the kaleidoscope literature
of fractional equations. The  fractional space derivatives  stem from  long trapping times while previously they were
interpreted as  a consequence of spatial
L{\'e}vy flights.
\end{abstract}
\maketitle

Fractional calculus is an old branch of mathematics that studies non-integer
differential operators \cite{Oldham1974Fractional,Samko1993Fractional,Podlubny1999Fractional,Meerschaert2012Stochastic}. This method is used extensively
to  model anomalous diffusion and relaxation in a wide variety
of systems \cite{Caputo,SchneiderWyss,MetzKlaf,Sokolov}.
 To recap consider the fractional
diffusion equation \cite{Saichev,Mainardi}
for the density of spreading particles ${\cal P}(x,t)$
\begin{equation}
\frac{ \partial^\beta {\cal P}} { \partial t^\beta} = D_{\beta,\mu} {\nabla^\mu}  {\cal P},
\label{eq01}
\end{equation}
where $D_{\beta,\mu}$, with units ${\rm m}^\mu/ \mbox{s}^\beta$,
is a generalised diffusion constant.
The fractional time and space
 derivatives are  convolution operators
that more intuitively are  defined with
their respective Laplace and Fourier transforms (see below). This equation, sometimes called the fractional diffusion-wave equation, reduces to the diffusion equation
when $\beta=1,\mu=2$
and the wave equation for $\beta=2,\mu=2$. $\mu<2$ corresponds to long spatial jumps referred to as L\'evy flights (LF), while $\beta<1$ to long dwelling times between jump events \cite{MetzKlaf}.  Originally this equation was derived using
the continuous time random walk (CTRW)  model \cite{Kotulski1995Asymptotic,MetzKlaf,Scalas,Carioli,Kutner,Morales2017Stochastic,Burioni2014Scaling}. More recently the
fractional diffusion equation with $\beta=1$
was  derived for heat transport using models of
interacting particles  \cite{Kundu,Dhar}.
Such fractional kinetic equations are
 used to describe the time of flight experiments
of charge carriers in
disordered systems where due to trapping
$\beta<1, \mu=2$
\cite{Scher75,fracPRE}
and anomalous  diffusion of cold atoms
in optical lattices where the atom-laser interaction induces  $\mu<2$ and $\beta=1$ \cite{Sagi,KesBarPRL}. Extensions that include  external forces are  well studied, within a framework referred to as
the  fractional Fokker-Planck equation \cite{FracPRL,Marcin,Deng2008Finite,Henry1},
and distributed order fractional equations \cite{Chechkin2005Fractional,Fedotov}.
 For an extensive review see
\cite{MetzKlaf}.


Eq. (\ref{eq01}) exhibits reflection symmetry and hence the packet of spreading  particles is  symmetric around its mean, if the initial condition density is localised.
In disordered systems with fixed advection,
symmetry breaking is found, and Eq.~\eqref{eq01} is invalid.
 Such  behavior is found  throughout hydrology, for example,
for tracer and contaminant  spreading  in heterogeneous media.
For more than two decades, two opposing and competing frameworks were developed  in this field.
One approach advanced
by  Benson, Schumer, Meerschaert, and
Wheatcraft (BSMW) \cite{Benson,Shumer}
proposed  that  the mechanism for transport is controlled by
non-local spatial jumps
of the L\'evy type \cite{MetzKlaf,Zhang,Kelly,zhang2005mass}.
It was suggested  that solute particles may experience long movements in
high velocity flow paths, leading to such super diffusive
behavior, possibly in the spirit
of LFs in rotational flow \cite{Swinney}.
 Importantly, since field observations
exhibit non-symmetric shapes of the spreading packet of particles,
the microscopic picture introduces skewed  probability density function
(PDF)  of spatial jump
lengths.
This approach  extensively promoted the use of non-symmetrical
fractional space advection-diffusion equations for LFs, see \cite{Kelly} for an overview.

The second approach uses what might be considered
the opposite strategy.
Instead of long non-local L{\'e}vy jumps in space,
Berkowitz, Scher and co-workers \cite{Scher,Berkowitz2002,Margolin,Levy,Berkowitz,Edery,Nissan,Morales2017Stochastic}
 showed that the  CTRW framework with a power-law
trapping time PDF is the key feature needed to explain the
observed data.
 Physically this is  the result of long trapping
events in geometrically induced  dead-ends
found in strongly disordered  porous media.
Specifically, based on field experiments and extensive  modelling,
the trapping time PDF is  $\psi(\tau) \sim \tau^{-(1+\beta)}$ and importantly in
many cases  $1<\beta<2$ \cite{Levy,Berkowitz}.
Here the mean trapping time is  finite, while the variance diverges.
In this case Eq. (\ref{eq01})  is certainly not valid. To see this
consider a CTRW with a finite variance of jump lengths,
so $\mu=2$  and then as  mentioned
if we take $ \beta \to  2$ we get the wave equation,
which is completely irrelevant for the transport under study.
Thus, so far, the  Scher-Berkowitz theoretical framework is based on a random walk picture \cite{Berkowitz2002} and
not on a governing fractional advection-diffusion equation. 
 Both the CTRW and the BSMW frameworks and
the experiments in the field agree on one
thing: advection-diffusion is anomalous and non-symmetric \cite{Burioni2013Rare,Berkowitz,Hou,WanliPRR}, however
otherwise these schools promote widely different philosophies.

One  goal of
this letter is to promote a better understanding of the
meaning of the fractional space derivatives in transport
equations. As mentioned in the literature these are associated
with LFs however we show here that they are
actually related to the long-tailed PDF of trapping times,
provided that $1<\beta <2$. 
The mentioned  biased CTRW  is known to exhibit super-diffusion $\langle (X - \langle X\rangle)^2\rangle \propto t^{3-\beta}$ \cite{Shlesinger1974Asymptotic} however
this as a stand alone does not imply a connection to LFs or fractional space
kinetic equations. 
The first important conceptual step towards  a unification of LFs and biased CTRW was given by Weeks, Urbach, and Swinney \cite{Weeks1998Anomalous,Weeks1996Anomalous}.
In the presence of bias, an observer in a reference frame moving with the mean speed set by the advection, will view the power-law trapping times of the CTRW framework, as if the particle is performing large jumps in space.
Our challenge is three fold. First to extend this idea into a fractional equation showing  the role of fluctuations. Secondly, to develop a tool, capable of dealing with a wide variety of  applied problems, ranging  from calculations of breakthrough curves (see below), effect
of time-dependent fields (omnipresent in field experiments) and different boundary  conditions by far extending  \cite{Burioni2013Rare,Hou,WanliPRR,Shlesinger1974Asymptotic,Weeks1998Anomalous,Weeks1996Anomalous}.  In essence
this framework  is the continuum fractional  diffusive description of a very large class of random walk processes. Finally, after deriving the fractional equation for the Berkowitz-Scher transport, we will be in the position to compare
it to the BSMW LF method.

{\em The fractional advection-diffusion-asymmetry  equation} (FADAE)
investigated in this letter reads
\begin{equation}
\frac{\partial}{\partial t} {\cal P} = D {\partial^2 \over \partial x^2 } {\cal P} - V  {\partial \over \partial x} {\cal P}  + S {{\rm \partial}^{\beta}  \over {\rm \partial} (-x)^\beta}  {\cal P}.
\label{eq02}
\end{equation}
The first two terms on the right-hand-side of Eq.~\eqref{eq02} are the standard
diffusion and drift terms, the last term is the modification we propose.
 The operator $ {\rm \partial}^\beta/ {\rm \partial} (-x)^\beta$ is a  Riemann-Liouville  fractional derivatives \cite{Samko1993Fractional,MetzKlaf} of order $1<\beta<2$; see the Appendix.
The Fourier transform of this operator acting on some test function is
${\cal F} \left[ {\rm d}^\beta  g(x) / {\rm d} (-x)^\beta \right] = (-ik)^\beta \tilde{g}(k)$ where $\tilde{g}(k)$ is the Fourier transform of $g(x)$.
In contrast to the spatial Riemann-Liouville derivatives in Eq. (\ref{eq02}),
 the generalised Laplacian
in Eq. (\ref{eq01}) is  symmetric Riesz derivatives
 \cite{Saichev}, where ${\cal F} [ \nabla^\mu g(x)] = - |k|^\mu \widetilde{g}(k)$.
Further in Eq. (\ref{eq02})
 we have no fractional time derivatives and hence obviously it is very
different from the standard fractional diffusion Eq. (\ref{eq01}). 
Here  $D$ describes  normal diffusion, $V$ controls the drift, while $S$ is the symmetry breaking parameter. We now explain the meaning of Eq.~\eqref{eq02} and its extensions.

When initially  ${\cal P} (x,t)|_{t=0}=\delta(x)$, namely  the packet of particles
is localised on the origin, and when the transport coefficients are time-independent and for free boundary conditions, the solution is obtained using Fourier transform.
Let $\tilde{{\cal P}}(k,t)$ be the Fourier pair of ${\cal P} (x,t)$ then
Eq.
(\ref{eq02}) gives
\begin{equation}
\tilde{P}(k,t) = \exp\left[ -  D k^2 t - i k V t + S (-i k)^\beta t\right].
\label{eq03}
\end{equation}
Thus the solution is a convolution
 of a Gaussian and a non-symmetric  L\'evy density \cite{Schneider1986Stable,Zolotarev1986One,Feller1971introduction,Uchaikin2011Chance,Padash2019First}.
These correspond
to limit distributions of sums of independent identically distributed random variables described by thin and  fat-tailed densities respectively. More specifically we denote $L_\beta(y)$ as the asymmetrical L\'evy density 
whose Fourier transform is   $\exp[ (-i k)^\beta]$, and hence
${\cal P} (x,t) = L_{\beta} [x/ (S t)^\beta](S t)^{-\beta}\bigotimes
\exp[ - (x - V t)^2 /4 D t]/\sqrt{4 \pi D t}$
where $\bigotimes$ is the convolution symbol \cite{WeylD,Otiniano2013Stable}. 

 {\em Model.} We treat the problem using the assumption that the particle will
wait for some random time $\tau$ between two successive jumps. This is exactly
the framework of the CTRW that describes a particle performing random independent
steps $x$, determined by the PDF $f(x)$, and the waiting time $\tau$ distributed according to $\psi(\tau)$
\cite{Kotulski1995Asymptotic,MetzKlaf,Hou,WanliPRR,Shlesinger1974Asymptotic,Weeks1998Anomalous,Weeks1996Anomalous}.
All the waiting times and the jump lengths are independent. 
We consider, $\psi(\tau) \sim \beta  (\tau_0)^{\beta}  \tau^{-1-\beta}$
and as mentioned $1<\beta<2$. 
The time scale $\tau_0$ together with the finite mean waiting time $\langle \tau \rangle =
\int_0 ^\infty \tau \psi(\tau) {\rm d} \tau$ is important.
The probability of observing $N$ steps at time $t$ is \cite{Godreche,WanliPRE}
\begin{equation}
Q_t (N) \sim \frac{1}{(t/\bar{t} )^{1/\beta}}
L_{\beta} \left[\frac{N - t/\langle \tau \rangle}{(t/\bar{t})^{1/\beta}}\right]
\label{eq04}
\end{equation}
with $\bar{t} = \langle \tau \rangle^{1 +\beta} /[(\tau_0)^\beta |\Gamma(1-\beta)|] $.
This equation  is valid for large times and large $N$, for example
 the mean number
of jumps $\langle N \rangle \sim t/\langle \tau \rangle$ is large.
 Eq. (\ref{eq04})  means that L\'evy
statistics are applicable for the shifted observable $N-\langle N\rangle$.
For the jump length distribution $f(x)$, we assume that the mean size
of the jumps is $a$ and the variance is $\sigma$. For example,
in simulations below, the PDF of jump size is Gaussian
\begin{equation}
f(x) =\frac{1}{\sqrt{2 \pi \sigma^2}}\exp \left[- \frac{(x -a)^2}{2 \sigma^2 }\right].
\label{eq05}
\end{equation}
%
%
The parameter $a$ is the bias, and the mean position of the particle after $N$
steps, is $N a$ hence on average the packet of particles starting on the
origin will be on $a t /\langle \tau \rangle$.
Clearly this modelling implies that we do not assume fat-tailed jump
length distributions, unlike the LF picture in BSMW \cite{WeylD}.

 In CTRW the position of the
 particle after $N$ steps is $X = \sum_{i=1} ^N x_i$,
and thus it depends both on the microscopic displacements $x_i$ and
the random number of  steps $N$. By conditioning
on a specific outcome of $N$ displacements, the PDF of finding the particle
at $X$ at time $t$ is
%
${\cal P}_{{\rm CTRW} }  (X,t) = \sum_{N=0} ^\infty Q_t (N) P(X|N)$.
%
We are interested in the long time limit since in this limit $N$ is large,
hence we replace $P(X|N)$ with the Gaussian,
and similarly replace $Q_t(N)$ with the L\'evy distribution Eq. (\ref{eq04}). Switching from summation to integration, in the long time limit we find
\begin{equation}
{\cal P}_{{\rm CTRW}} (X,t)  \sim 
\int_{0} ^\infty L_{\beta} \left( { N - t/\langle \tau \rangle \over (t /\bar{t} )^{1/\beta} }\right){ \exp \left( - { (X- a N)^2 \over 2 \sigma^2 N} \right) \over \sqrt{ 2 \pi \sigma^2 N}(t / \bar{t} )^{1/\beta}} {\rm d} N.
\label{eq07}
\end{equation}
This idea  is also known as the subordination of the spatial process
$X$ by the temporal process for $N$ and is routinely considered
in the literature for $\beta<1$, see  \cite{Fogedby,fracPRE}.
We already mentioned our intention to derive the spatial derivative  usually
associated with L\'evy spatial jumps using the perfectly Gaussian jump statistics in space,
 and that is what we do next. In the long time limit, we find
\begin{equation}
{\cal P}_{{\rm CTRW}} \sim \int_{-\infty} ^\infty L_{\beta} (y) {\exp \left\{  - {   \left[X - a t /\langle \tau \rangle- a y (t/\bar{t})^{1/\beta} \right]^2
\over 2 \sigma^2 t/\langle \tau \rangle}  \right\} \over
\sqrt{ 2 \pi \sigma^2  t /\langle \tau \rangle } } {\rm d} y .
\label{eq09}
\end{equation}
Technically this limit is obtained with a change of variables to
 $\xi= (X - a t/\langle \tau \rangle) /a (t/\bar{t})^{1/\beta}$
and $\xi$ is kept fixed while $t \to \infty$ \cite{WeylD}.
We now take the time derivative
of the Fourier transform of Eq. (\ref{eq09}) and find
\begin{equation}
{\partial \tilde{{\cal P}} (k,t) \over \partial t} =-
{ \sigma^2 \over 2 \langle \tau \rangle} k^2 \tilde{{\cal P}} (k,t) - i k {a \over \langle \tau \rangle} \tilde{{\cal P}}(k,t) + (-i k)^\beta {a^\beta \over \bar{t}} \tilde{{\cal P}} (k,t).
\label{eq10}
\end{equation}
This is the Fourier representation of Eq. (\ref{eq02})
 when we  identify the
transport constants:
\begin{equation}
D = {\sigma^2 \over 2 \langle \tau \rangle},\  V = {a \over \langle \tau \rangle},\
S= {a^\beta \over \bar{t} } .
\label{eq11}
\end{equation}
The two formulas for $D$ and $V$
are standard relations
in the theory of advection-diffusion.
To summarize, the FADAE (\ref{eq02})
describes
the biased CTRW process and this has several consequences which are
now discussed.

\begin{figure}[htb]
 \centering
 \includegraphics[width=1.0\textwidth]{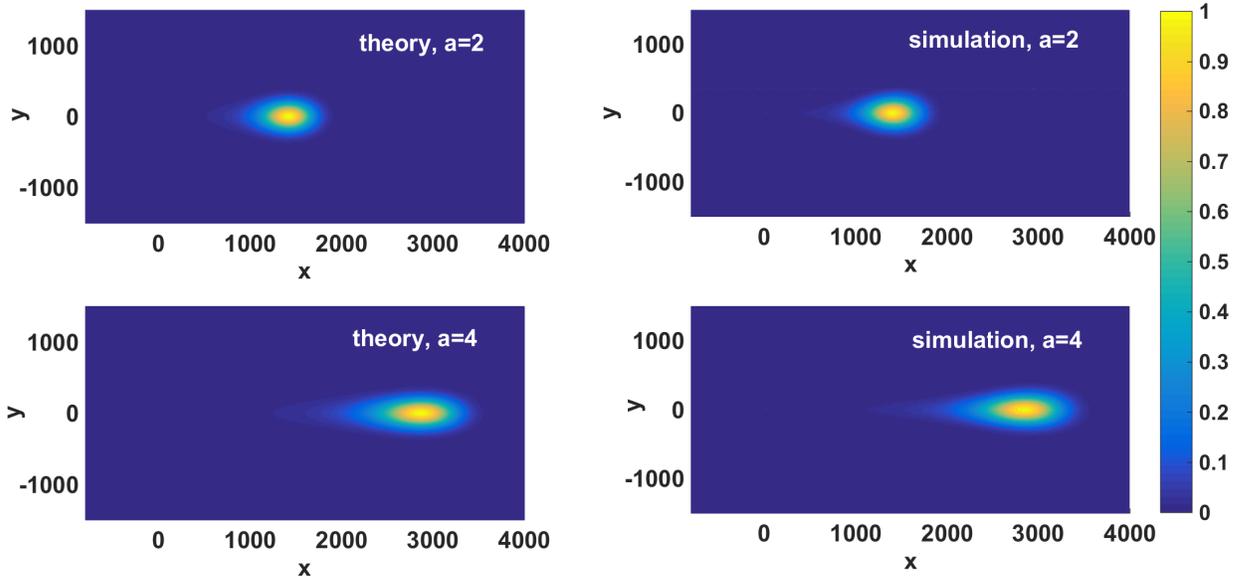}\\
 \caption{Packets of particles released from an origin in two-dimensions  with $\beta=3/2$, time
$t=200$ where the mean waiting time is
 $\langle \tau\rangle=0.3$ and $\tau_0=0.1$.
 The bias is pointing to the $x$ direction, while it is absent along the $y$
axis, and this creates packets distorted in the direction of the field. The symmetry breaking
 effect
is visibly stronger as the bias level is increased.
Here we show how simulations of the CTRW process and the analytical solutions of the FADAE nicely match. For theory we use
Eq. (\ref{eq11}) which gives
 $D=41.7,V=a/0.3$, and $S=a^{3/2}/0.44$ the bias $a$ is provided in the figure while in the $y$ direction $D=41.7$, and $S=V=0$.
For further details on simulations
see the Appendix, for example,  a perfect agreement between theory and simulations without fitting, for one-dimensional  CTRW. 
}\label{fig1}
\end{figure}

{\em The importance of bias.}
An interesting effect is that in the absence of bias, i.e. $a=0$
 we get $S=0$,
hence the anomaly is present only when we have advection. Since $S=0$
implies normal diffusion, in the case of weak advection the solution
exhibits nearly normal behaviour even for very long times, an effect
 crucial
for experiments.
Further,  Eq. (\ref{eq11})
 shows how the
two transport coefficients $S$ and $V$  are generally  not independent.
To see this consider
linear response theory.
Then we have $a \sim F$ where $F$ is the external force field, and
we have $V\sim F$ and $S\sim F^{\beta}$, a prediction that could be tested
in experiments.

{\em Packets in  two dimensions.} The fact that the asymmetry constant $S$
 is bias-dependent leads to the following interesting prediction
in   two dimensions. Imagine the bias is directed in the $x$ direction,
then the distortion of the packet of particles is found only along the $x$
axis.
In other words the diffusion in the perpendicular
 $y$ direction will be perfectly normal.
In the Appendix  we extend our mathematical  treatment of the problem to two dimensions. Here we present this effect graphically in  Fig. \ref{fig1},
 the asymmetrical  oval
like
shape of the spreading  packet is clearly visible, with the left tail broader than the
right one. Similar experimental  observations were reported
in \cite{Levy,Berkowitz2009shcer}.
The left tail, seen clearly in the figure, is due to trapping of particles
far lagging behind the mean position of  the packet and this as we
showed is modeled with the asymmetry operator $\partial^\beta
 / \partial (-x)^\beta$   in Eq. (\ref{eq02}).
Thus the physical interpretation of the fractional space derivatives
 in FADAE should be made with care,
as it does not necessarily
mean that the process exhibits LFs.

\begin{figure}[htb]
 \centering
 \includegraphics[width=1.0\textwidth]{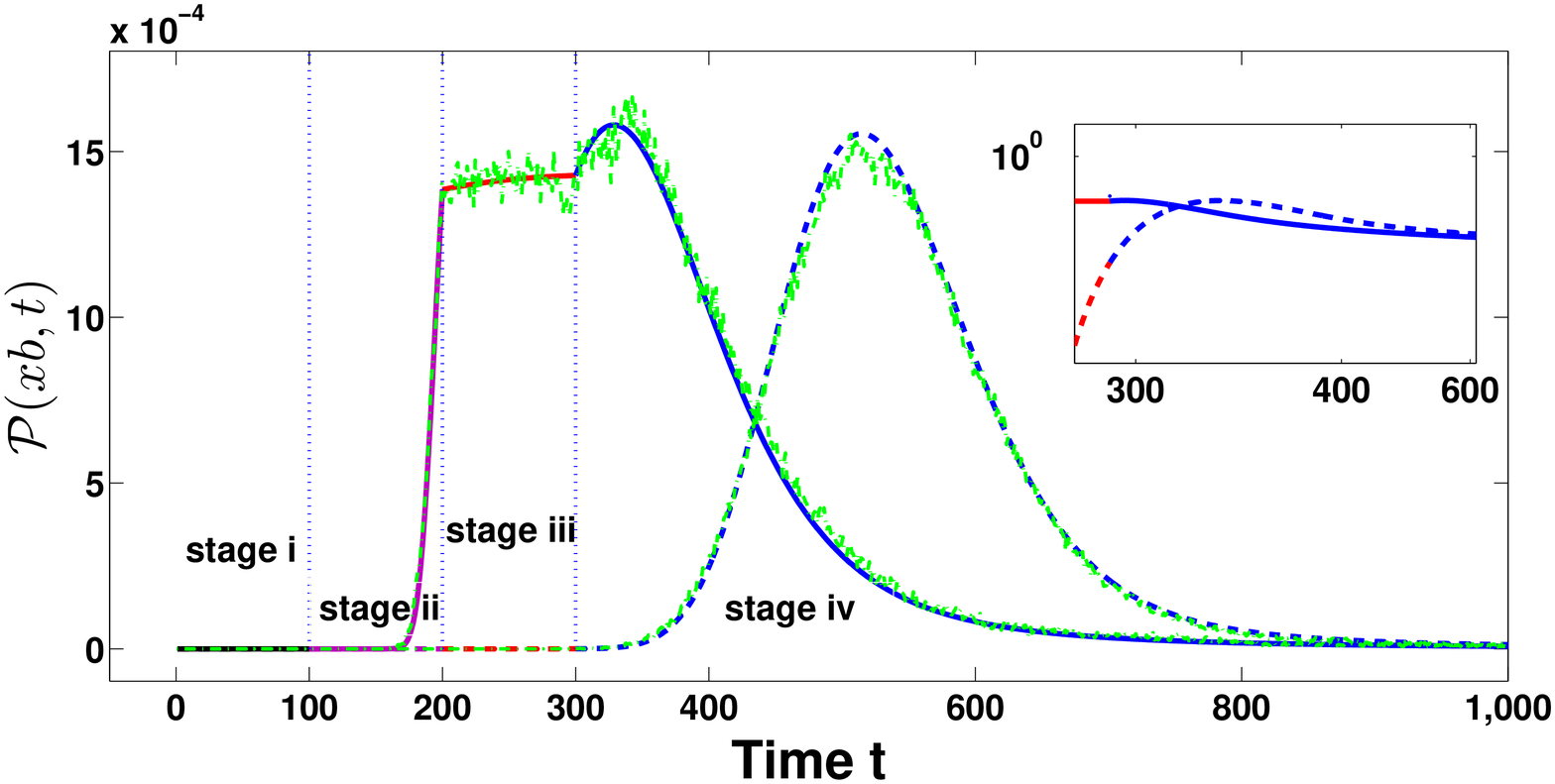}\\
 \caption{Particles  are  released on the origin at time zero
 and then the density on $x_b$ is
recorded versus time. Such breakthrough curves  present contaminant
spreading from a source (say upper part  of a stream) to some target on $x_b$, here $x_b=1800$.
Here we present the solution of the FADAE and compare it with the CTRW simulations with $\beta=3/2$ without fitting.
 The bias $a$ is
time-dependent, and as explained in the text, the dynamics has four stages
as indicated in the graph, e.g. stage i), $0<t<100$ etc.
}\label{fig2}
\end{figure}

{\em Temporal variations} of the mean velocity $a/\langle \tau \rangle$ is often present in the real world and tested experimentally in \cite{Nissan}. We explore
this issue now using a time-dependant but piece wise constant bias $a(t)$
 \cite{Nissan}. Indeed  in controlled experiments, the velocity $V$ can be modified, and then theoretical predictions can be tested in a non trivial
setting.  This  example will demonstrate
the power  of the fractional framework, as it allows for a semi-analytical solution  of the  rather complex
behaviour, and present physical effects related to the magnitude of the bias.
We consider four stages of the transport \cite{Nissan}:
 i) we use bias $a=1$  ii) we then sharply increase $a$ to a value $a=3.6$,
then (iii)
 decrease the value then bias  to a small number $a=0.09$, and finally (iv)
 return to the bias in  state i). All along the second length scale
$\sigma=5$  is fixed.
The time lapses of each stage clearly indicated
in
Fig. \ref{fig2} while
the derivation of analytical results is left to the Appendix.
Note that as we modify the bias $a$ we are effectively modifying $V$ and $S$
while $D$ remains fixed, see Eq.  \eqref{eq11}.
 The essential idea behind the analytical approach
 is that the final state  of
each stage  serves as an initial condition to the spatial distribution
of the  next stage.
In Fig. \ref{fig2}  (curve A)
this analytical method is compared
to numerical solution of the CTRW with $\beta=3/2$, finding excellent agreement.
We also present  the case of a constant time-independent  $a=1$ (curve B). The concentration $P(x_b,t)$ at some fixed $x_b$ presented in Fig. \ref{fig2}, is called a breakthrough curve and it is commonly observed in the field  of contaminant spreading in
 Hydrology. 
 Fig. \ref{fig2}  clearly demonstrates the excellent
 quantitive agreement between theory
and simulation, in a regime of dynamics which is close
to real real life experiments and far from trivial.
Hence we are confident that our tool, the FADAE is a useful one.

{\em L\'evy flights and the interpretation of  experiment.}
The CTRW process with long-tailed PDFs is  an excellent model
for transport in a wide variety of systems, for example porous media,
 hence the governing FADAE (\ref{eq02}) is deeply related to transport
in many
physical systems \cite{MetzKlaf,Kutner,Hou,Scher75,Bouchaud,Andrea,Scher2,PCCP}.
 Still it is interesting to compare our approach
to the fractional model of LFs that reads \cite{Benson,remark}
\begin{equation}
{\partial{\cal  P}_{{\rm LF}} \over  \partial t} =- V {\partial {\cal P_{{\rm LF}}} \over \partial x} +  K \left( q {\partial^\mu  \over \partial (-x)^\mu } + p {\partial^\mu \over \partial x^\mu} \right) {\cal P}_{{\rm LF}}
\label{eq12}
\end{equation}
Clearly this equation is very different from ours, in fact in some sense it  is more general as compared with  Eq. \eqref{eq02}, as it describes a general class of skewed processes with the phenomenological  parameters $p$ and $q$.  In \cite{zhang2005mass},
authors fit experimental
of contaminant
data and report: $V=0.8~{\rm m/h}$, $D=0$, $\mu=1.51$, $K=2.8~{\rm m^{1.51}/h}$, $q=1$, and $p=0$ to match the breakthrough curves.  Based on this one may naively interpret the data as stemming from a LF process. However, we realise that these parameters imply, based on our notation Eq.~\eqref{eq02}, a strong bias in the long time limit of CTRW.  This highlights that the data are consistent with a CTRW with long-tailed trapping times. To summarize, using $p=0$ in Eq.~\eqref{eq12} is consistent  with both a LF picture promoted by BSMW and a CTRW with broad-tailed  waiting times. 

To distinguish between  these  two approaches one needs to analyse the trajectories  of the process, not the packet of the spreading particles. More exactly, the CTRW approach and LFs method can give the same predictions for the positional distribution, but the interpretation that a model with fractional space derivatives always implies LFs is wrong.
In that sense we claim that the two competing methods are identical (in some limit relevant to experiments) from the point of view of distributions but the particles trajectories  widely differ.

{\em Extensions with subordination}.
 A key formula  is the transformation Eq. (\ref{eq07}). It shows how
to transform a normal process to an anomalous one,  for  the case
$1<\beta<2$ and as mentioned, this idea is called subordination.
In Eq. (\ref{eq07})  t is the laboratory time
and $N$ is sometimes referred to as operational time.
The idea is simple, $N$ which is actually the random number
of steps in the process, is distributed according to L\'evy statistics,
as expected from the generalised central limit theorem.
 We then transform the Gaussian process in the
operational time $N$  to
the laboratory framework with what we call a
L\'evy transformation, see Eq.  (\ref{eq07}).
This method, can be extended to include cases with different boundary conditions, different spatial dependent  force fields, stochastic trajectories
etc.
and hence the mathematical  approach we presented  is versatile
and  far more
general than what we considered here.

{\em Mean square displacement.} All along the manuscript
we focus on the typical fluctuations of the process. The rare events influence the density $\mathcal{P}(x,t)$ in the vicinity of $x\simeq0$ \cite{WanliPRR}. Here  Eq.~\eqref{eq02}  does not work.
In that limit,
the probability of finding these particles is very small, as expected for a biased process.
In fact such a cutoff exists also for the diffusion equation, where the telegraph
equation  can be used to describe  far tails of the density of particles.  However, in the present case rare events control the behavior of the mean square displacement, which exhibits superdiffusion.  It indicates that Eq.~\eqref{eq02} cannot give a valid mean square displacement \cite{Hou,WanliPRR}, in this sense CTRWs are of course very different compared to LFs.

{\em Summary.}  The FADAE (\ref{eq02}) is controlled by three transport
coefficients, $D$, $V$ and $S$, given in Eq. (\ref{eq11}).
This framework is valuable in many CTRW  systems,
 ranging from the  field of contaminant
spreading and geophysics
to transport random environments,
for example, the quenched trap model \cite{Bouchaud,Burov}.
What is remarkable is that the long-tailed  PDF of trapping times, which for
$0<\beta<1$  implies a fractional time derivative,
is transplanted into a
spatial space derivative when $1<\beta<2$.
 And
 long tailed PDFs of jump sizes, like in LFs, are not a
basic  requirement
for fractional space  operators in transport  equations,
rather these are related to L\'evy statistics applied to the number of jumps
in the process. In this sense we have provided a new physical and widely applicable
interpretation   of
fractional space derivatives, within   the context of fractional
diffusion. More importantly, we provided a tool box with which one may analyse advection diffusion with different boundary conditions (with subordination) and with time-dependent fields.

\begin{acknowledgments}
E.B. acknowledges the financial  support of the Israel Science Foundation
grant 1898/17.
\end{acknowledgments}
\appendix
\begin{appendices}

\section{Mathematical background}
Here we briefly discuss fractional derivatives, the non-symmetric stable probability density function, Eq. (7) in the main text, the infinite density that controls the rare fluctuations of $X$,  the convolution used in the main text, and the general fractional advection-diffusion-asymmetry equation. 
\subsection{Riemann-Liouville derivatives}
There are many excellent texts describing the long history and
analytical properties of fractional derivatives, for example see Refs. \cite{Oldham1974Fractional,Podlubny1999Fractional,Meerschaert2012Stochastic}. The fractional space derivatives operator $\frac{{\rm d}^\beta}{{\rm d} (-x)^\beta}$ used in the main text is now discussed. Here we  focus on Riemann-Liouville  derivatives \cite{Oldham1974Fractional,remark}. In real space, the (right) Riemann-Liouville  derivative operator is defined through
\begin{equation}\label{Weyl1}
\frac{{\rm d}^\beta}{{\rm d} (-x)^\beta}g(x) =\frac{(-1)^n}{\Gamma(n-\beta)} \frac{{\rm d}^n}{{\rm d} x^n}\int_x^\infty (y-x)^{n-\beta-1}g(y)\rm{d}y,
\end{equation}
where $n$ is the smallest integer  larger than $\beta$. While if the bottom limit of the integral is set to minus infinity, we have  another related expression called left Riemann-Liouville  derivatives
\begin{equation}\label{Weyl2}
\frac{{\rm d}^\beta}{{\rm d} x^\beta}g(x) =\frac{1}{\Gamma(n-\beta)} \frac{{\rm d}^n}{{\rm d} x^n}\int_{-\infty}^x (x-y)^{n-\beta-1}g(y)\rm{d}y.
\end{equation}
The mentioned operators have a simple expression under transforms since no initial values come into play.
In Fourier space, the Riemann-Liouville derivative derivatives obey the following theorem \cite{Oldham1974Fractional,remark}
\begin{equation}\label{weyl3}
\left\{
  \begin{array}{ll}
 \displaystyle \mathcal{F}\left[\frac{{\rm d}^\beta}{{\rm d} (-x)^\beta}g(x)\right]=(-ik)^\beta \tilde{g}(k), & \hbox{} \\
 \displaystyle \mathcal{F}\left[\frac{{\rm d}^\beta}{{\rm d} x^\beta}g(x)\right]=(ik)^\beta \tilde{g}(k), & \hbox{}
 \end{array}
\right.
\end{equation}
where we denote $\tilde{g}(k)$, $\tilde{g}(k)=\int_{-\infty}^\infty\exp(-ikx)g(x){\rm d}x$, as the Fourier transform  of $g(x)$. 

\begin{figure}[htb]
  \centering
  \includegraphics[width=13cm]{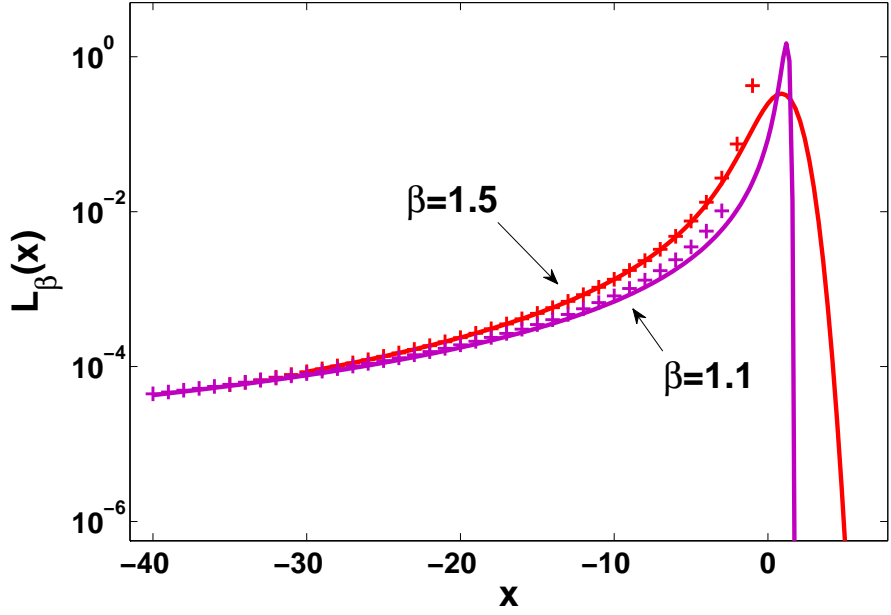}\\
  \caption{Plot of the L{\'e}vy distribution with $\beta=3/2~ {\rm and}~\beta=1.1$, showing asymmetric far tails. The L{\'e}vy distribution is tabulated with Mathematica (see the solid lines)  and the asymptotic behavior of the left  tail, plotted by the symbols `$+$', is obtained from  Eq.~\eqref{lefttail} [the last line].  }
\label{LevyDistribution}
\end{figure}

\subsection{Non-symmetric L{\'e}vy distribution}
As mentioned in the main text, in the long time limit, the number of jumps $N$ in the renewal process, shifted by its mean $\langle N\rangle$, follows the L{\'e}vy stable distribution, which is defined by
\begin{equation}\label{Levydistribuion}
L_{\beta}(x)=\frac{1}{2\pi}\int_{-\infty}^\infty \exp(ikx)\exp[(-ik)^\beta]{\rm d}k.
\end{equation}
If $0<\beta<1$, the above equation reduces to the one-sided L{\'e}vy distribution, namely $L_\beta(x)=0$ for $x>0$. 
In the literature, some authors prefer to use the characteristic function to define the L{\'e}vy distribution, for example, see Refs. \cite{Schneider1986Stable,Zolotarev1986One,Feller1971introduction,Padash2019First}, while we use the Fourier transform (the difference is simply a sign in front of $k$).
Eq.~\eqref{Levydistribuion} is plotted in Fig.~\ref{LevyDistribution} using Mathematica command, i.e., ${\rm PDF}{\rm [StableDistribution[}1, \beta, -1, 0, {\rm Abs[Cos[Pi*}\beta/2]]^{1/\beta}], x]$; see also the inset in linear-linear scale. Rewriting  Eq.~\eqref{Levydistribuion}, we have
\begin{equation}
L_{\beta}(x)=\frac{1}{\pi}\int_0^\infty \exp\left(k^\beta\cos\left(\frac{\pi\beta}{2}\right)\right)\cos\left(kx-k^\beta\sin\left(\frac{\pi\beta}{2}\right)\right){\rm d}k,
\end{equation}
which can be used to plot the PDF of the L{\'e}vy distribution.
We are  interested in the decay of the far left and  right tails, which are   briefly  presented below to  emphasize  the difference between them.

In the large deviation regime, namely $x$ is large, from Eq.~\eqref{Levydistribuion} the saddle-point method yields
\begin{equation}\label{righttail}
L_\beta(x)\sim \frac{\exp\left(-(\beta-1)(\frac{x}{\beta})^{\frac{\beta}{\beta-1}}\right)}{\sqrt{2\pi \beta^{-\frac{1}{\beta-1}}(\beta-1)x^{\frac{\beta}{\beta-1}}}},
\end{equation}
which was found originally in  Ref.~\cite{Zolotarev1986One} using other methods.
This indicates that the right tail of the distribution decays as $\exp(-(\beta-1)(x/\beta)^{\frac{\beta}{\beta-1}})$, approaching zero rapidly.

Splitting the integral Eq.~\eqref{Levydistribuion} into two parts at $k=0$ and changing variables $k^{'}=k$ for the negative $k$, we have
\begin{equation}\label{asy103}
L_{\beta}(x)=\mathcal{R}e\left[\frac{1}{\pi}\int_0^\infty\exp(-ikx)\exp((ik)^\beta){\rm d}k\right],
\end{equation}
where $\mathcal{R}e[g(x)]$ means the real part of $g(x)$. Expanding the integrand $\exp[(ik)^\beta]$ in the right hand side of Eq.~\eqref{asy103} as a Taylor series
\begin{equation}\label{asp104}
\exp[(ik)^\beta]=
\sum_{n=0}^\infty\frac{(-ik)^{\beta n}}{n!}=\sum_{n=0}^\infty \frac{1}{n!}(-ik)^{\frac{n+\frac{1}{\beta}-\frac{1}{\beta}}{\frac{1}{\beta}}},
\end{equation}
and substituting Eq.~ \eqref{asp104} into Eq.~ \eqref{asy103}, we get the asymptotic behavior of the left tail
\begin{equation}\label{lefttail}
\begin{split}
 L_\beta(x) &\sim \sum_{n=0}^\infty \mathcal{R}e\left[\exp\left(\frac{(n\beta+1)\pi}{2}\right)\Gamma(n\beta+1)\frac{i^{n\beta}}{n!(-x)^{n\beta+1}}\right]\\
    & =\sum_{n=1}^\infty-\frac{\sin(n\beta\pi)\Gamma(n\beta+1)}{\pi n!(-x)^{n\beta+1}}\\
    &\sim \frac{1}{\Gamma(-\beta) (-x)^{\beta+1}}
\end{split}
\end{equation}
with $x\to -\infty$.  The well known leading term, i.e., the last line of Eq.~\eqref{lefttail}, is plotted in Fig.~\ref{LevyDistribution} using the symbols `$+$'.
Note that here we used method of stationary phase  when calculating  Eq.~\eqref{lefttail} from the integral Eq.~\eqref{asy103}. As shown in Fig.~\ref{LevyDistribution}, we can see that two tails of the asymmetry L{\'e}vy distribution under study show different behaviors. The left one decays as a power law, i.e., $L_{\beta}(x)\sim (-x)^{-\beta-1}$ with $x\to -\infty$, tending to zero slowly if compared with the right one. This indicates that the variance of $x$ is infinity but the mean is finite.
The exact expression of L{\'e}vy distribution, exists in terms of Fox function,
which  can also be expressed in the form of Mellin-Barnes type of  integral. See Ref. \cite{Schneider1986Stable} for a review. There are a number of cases of analytically expressible stable distribution, for example, when $\beta=3/2$, the L{\'e}vy distribution is related to  Whittaker function \cite{Uchaikin2011Chance}.

\begin{figure}[htb]
 \centering
 \includegraphics[width=0.8\textwidth]{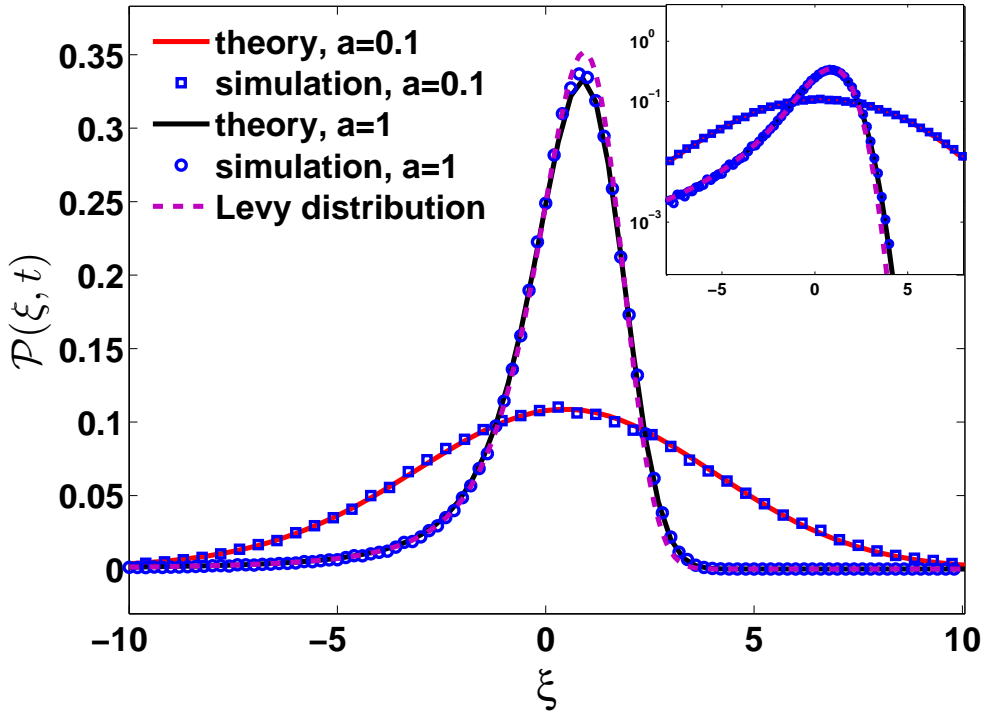}\\
 \caption{PDF of $\xi= (X - a t/\langle \tau \rangle) /a (t/\bar{t})^{1/\beta}$ for various $a$ listed in figure.  The  symbols denote the simulation results obtained by averaging $10^7$ particles with $\beta=1.5$, $\tau_0=0.1$, $\sigma=1$ and $t=1000$. The corresponding theoretical theory 
 plotted by solid  lines is obtained from  Eq.~(7) in the main text by changing variable showing excellent fitting. When the bias is weak, the  limiting law Eq.~\eqref{eq01s} plotted by the dashed line completely fails; see the simulations for $a=0.1$.
}\label{fig11}
\end{figure}

\subsection{Derivation of Eq.~(7)}
In the  Eq. (7) of the letter  we used a subordination method. We now explain a step in the derivation. According to subordination form Eq.~(6),  we change variables according to $y= (N - t/\langle \tau\rangle)/(t/\bar{t})^{1/\beta}$
and find
\begin{equation*}
 {\cal P}_{{\rm CTRW}}(X,t)  \sim
\int_{-\infty} ^\infty L_{\beta} \left( y \right)
{ \exp \left\{ - { \left[X- a t /\langle \tau \rangle - a y (t /\bar{t})^{1/\beta} \right]^2
 \over  2 \sigma^2 \left[ t/\langle \tau \rangle  + y (t /\bar{t})^{1/\beta}  \right] } \right\}
\over \sqrt{ 2 \sigma^2 \pi \left[ t/\langle \tau \rangle + y (t /\bar{t})^{1/\beta} \right] } }   {\rm d} y,
\label{eq08}
\end{equation*}
where the lower limit of the integration is found when $t\to \infty$.
The typical fluctuations of the process, are found when $X-at/\langle \tau \rangle \sim a (t/\bar{t})^{1/\beta}$,
and in this scaling regime, we finally find the law describing typical fluctuations, i.e., Eq.~(7).

\subsection{The Limit theorem and the infinite density.}
\begin{figure}[htbp]
\begin{minipage}[t]{0.49\linewidth}
\centering
\includegraphics[width=1\linewidth]{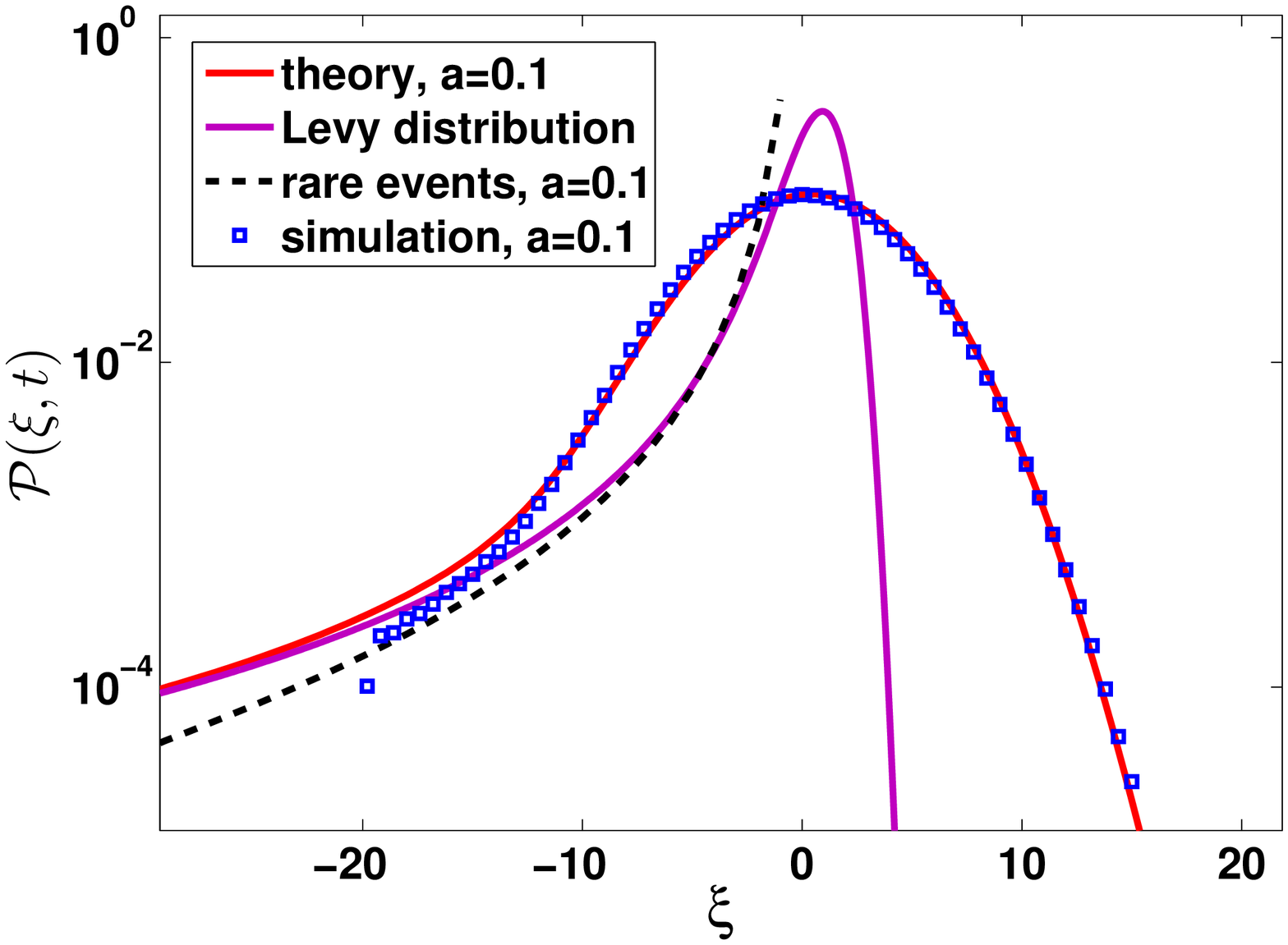}
\caption{Simulations of the distribution of $\xi$ with the scaling $\xi=(X - \langle X^*(t) \rangle)/l(t) $ and $\beta=1.5$ compared with the analytical
prediction  Eq. (7) in the main text. Here we choose $t=10^3$, $\tau_0=0.1$, $\sigma=1$, $a=0.1$ and $10^7$ trajectories for simulations.}\label{PxtVsAa01}
\end{minipage}%
\hfill
\begin{minipage}[t]{0.49\linewidth}
\centering
\includegraphics[width=1\linewidth]{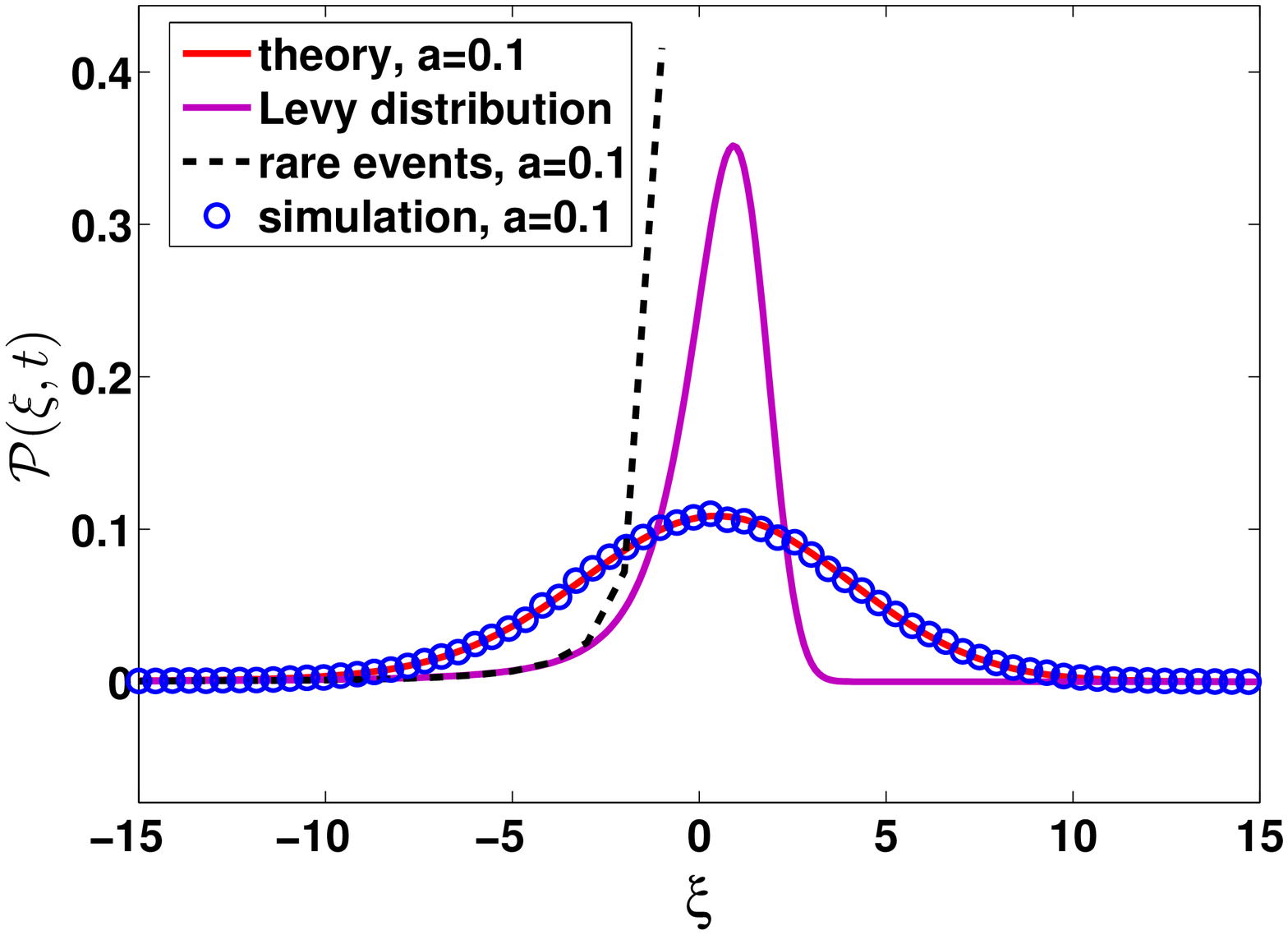}
\caption{Same as Fig. \ref{PxtVsAa01} in linear-linear scale.}
\label{PxtVsAa01linear}
\end{minipage}
\end{figure}

\begin{figure}[htbp]
\begin{minipage}[t]{0.49\linewidth}
\centering
\includegraphics[width=1\linewidth]{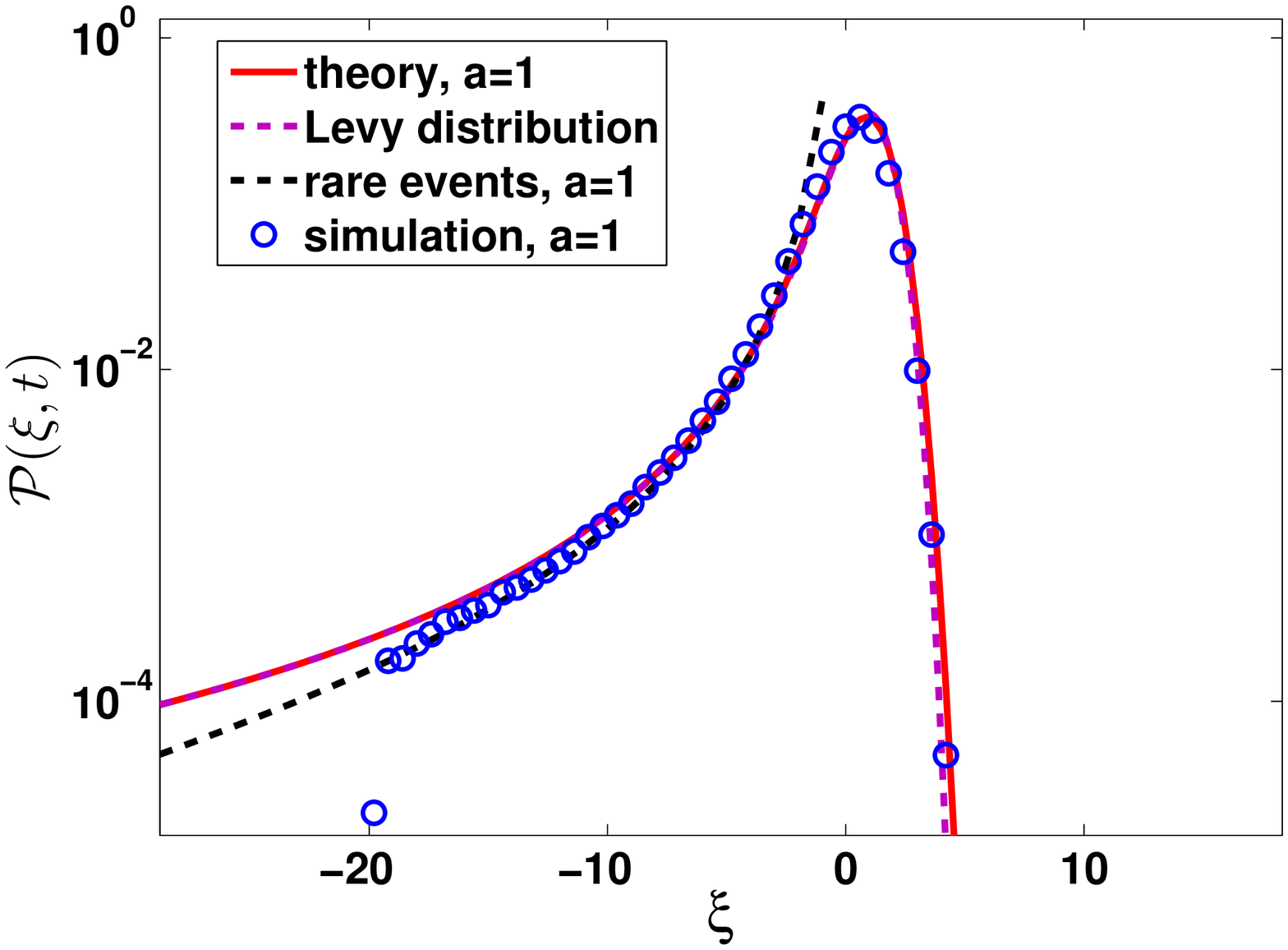}
\caption{Same as Fig.~\ref{PxtVsAa01} for $a=1$.}\label{PxtVsAa1}
\end{minipage}%
\hfill
\begin{minipage}[t]{0.49\linewidth}
\centering
\includegraphics[width=1\linewidth]{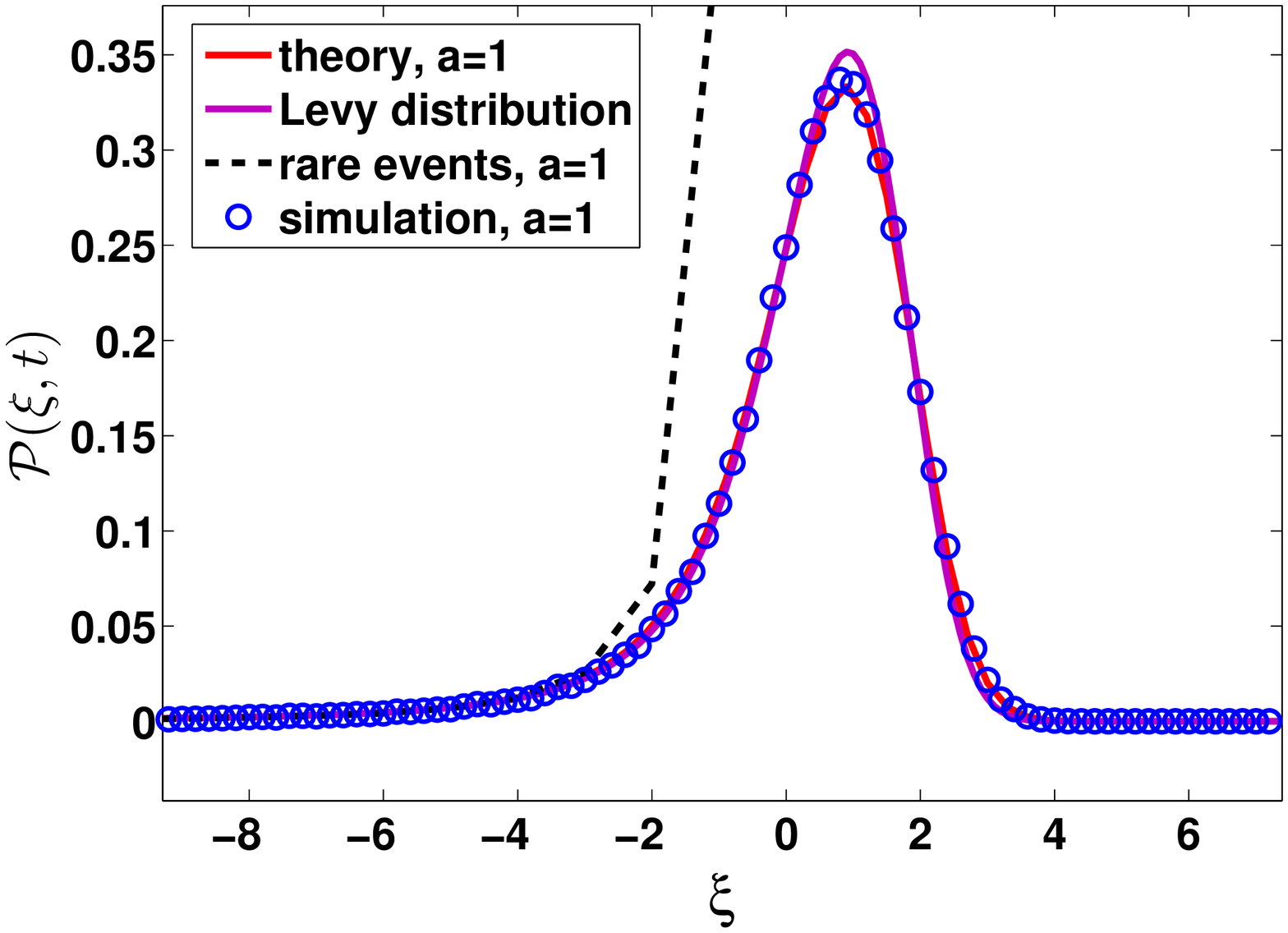}
\caption{Same as Fig.~\ref{PxtVsAa1}~in linear-linear scale.}
\label{PxtVsAa1linear}
\end{minipage}
\end{figure}

The mean position
 of the walker in the long time limit is clearly $\langle X(t) \rangle \sim
 a t /\langle \tau\rangle$,
 where $a$ is the mean displacement of each jump, and $t /\langle \tau \rangle$ is the average
number of jumps. For convenience assume that $a>0$, so $\langle X(t)\rangle$ is positive.  In the limit $t \to \infty$,
 it was rigorously shown that \cite{Kotulski1995Asymptotic,Burioni2014Scaling,WanliPRR}
\begin{equation}
P(X,t)_{{\rm CTRW}} \sim {1 \over l(t) }  L_{\beta} \left({ X - \langle X^*(t) \rangle
 \over l(t) } \right),
\label{eq01s}
\end{equation}
where $L_{\beta}(\cdot)$ represents the non-symmetrical L\'evy stable law whose  Fourier transform is $\exp[(-ik)^\beta]$ and $\langle X^*(t) \rangle$ is equal to $at/\langle\tau\rangle$. See the dashed line in Fig.~\eqref{fig11}.
Here $l(t) = a(t/\overline{t})^{1/\beta}$ describing the width of the distribution is
the typical length scale of the problem with $\overline{t} = \langle \tau\rangle^{1 + \beta} / (\tau_0)^\beta | \Gamma(1 - \beta)|$.
This equation while perfectly correct has several
drawbacks which we wish to address. Firstly, the left tail of  the density decays
as a power law, which is a well known property of   L\'evy distribution.
This naively
implies that the mean square displacement is infinite, which is certainly not possible. An immediate
consequence is that the mentioned super-diffusive effect discussed in the
recent literature,  is related to the non
typical fluctuations. Thus
it is not difficult to realize that this limit law has a cutoff, and this is solved recently in \cite{WanliPRR} 
\begin{equation}\label{19seq13}
\mathcal{P}(X,t)_{{\rm CTRW}}\sim \frac{(\tau_0)^\beta t^{-\beta}}{a}\mathcal{I}_{\beta}(\eta)
\end{equation}
with $0<\eta<1$, $\eta=1-(X/a)/(t/\langle\tau\rangle)$, and $$ \mathcal{I}_\beta(y)=\beta y^{-\beta-1}-(\beta-1)y^{-\beta}.$$ 
Eq.~\eqref{19seq13} is called the infinite density since the integration of Eq.~\eqref{19seq13} over $X$ diverges; see Figs.~\ref{PxtVsAa01}, \ref{PxtVsAa01linear}, \ref{PxtVsAa1}, and \ref{PxtVsAa1linear}.
Further discussion and simulations are presented in Ref. \cite{WanliPRR}.

Secondly, Eq. \eqref{eq01s} is independent of $\sigma$ which in
many physical  situations is not satisfactory. Let us consider a simple case, where the bias is weak and  $t$ is finite. We have $a\ll\sigma$ and then  expect that the variance of size of jumps is  a key parameter, for example when $a=10^{-2}$, $\sigma=100$, $t=100$, we expect to get the Gaussian distribution. Further, in the context
of  active rheology  and more generally single
molecule experiments the number of jumps $N$ can be large (say $100-1000$)
and the limit theorem which assumes  $\langle N\rangle\to \infty$
is found to be a non-sensible description
within the practical time range. As shown in Figs.~\ref{PxtVsAa01} and \ref{PxtVsAa01linear}, the limiting law  Eq.~\eqref{eq01s}  completely fails for small $a\ll \sigma$. With the growth of $a$, $P(\xi,t)$ approaches Eq.~\eqref{eq01s} slowly. Based on Eq.~\eqref{19seq13}, the MSD reads
\begin{equation}\label{eqsjs2}
\langle x^2(t)\rangle -\langle x(t)\rangle^2\sim \frac{2a^2(\beta-1)(\tau_0)^\beta |\Gamma(1-\beta)|}{\langle\tau\rangle^3\Gamma(4-\beta)}t^{3-\beta},
\end{equation}
showing enhanced diffusion. Note that Eq.~\eqref{eqsjs2}  can not be obtained from Eq.~(2) in the main text since we used the typical fluctuations of $N$. In particular, if we are not interested in the mean square displacement, the rare events can be ignored; see Figs. \ref{PxtVsAa01linear} and \ref{PxtVsAa1linear}.

\subsection{Convolution of L{\'e}vy and  Gaussian distribution}
Now let us consider  the solution of the fractional advection-diffusion equation with  an initial condition on the origin. As mentioned in the text the  solution $\mathcal{P}(x,t)$ in one dimension  is the convolution with respect to  L{\'e}vy  and Gaussian distributions
\begin{equation}\label{fadv101}
\mathcal{P}(x,t)=\int_{-\infty}^\infty \frac{1}{(St)^{1/\beta}}L_\beta\left(\frac{y}{(St)^{1/\beta}}\right)\exp\left(\frac{-(x-y-Vt)^2}{4Dt}\right)\frac{1}{\sqrt{4\pi Dt}}{\rm d}y.
\end{equation}
Such a convolution is sometimes called the Voigt profile, see related discussions in Ref. \cite{Otiniano2013Stable}.
The scaling behavior of $\xi=(x-Vt)/(St)^{1/\beta}$ gives
\begin{equation}\label{PDFXIiNT}
\mathcal{P}(\xi,t)=\int_{-\infty}^\infty L_\beta\left(z\right)\underbrace{\exp\left(\frac{-(\xi-z)^2}{4Dt (St)^{-2/\beta}}\right)\frac{1}{\sqrt{4Dt}(St)^{-1/\beta}}}_{~}{\rm d}z.
\end{equation}
Below we investigate two  limits:
\begin{itemize}
\item When $\mathcal{Z}=4Dt(St)^{-2/\beta}\to 0$, i.e., $4Dt\ll (St)^{2/\beta}$, the term marked by under-brace in Eq.~\eqref{PDFXIiNT} reduces to a delta function using $\lim_{\epsilon\to 0^+}\exp(-z^2/(4\epsilon))/\sqrt{4\pi\epsilon}=\delta(z)$. Thus in the very long time limit the L{\'e}vy distribution  describes the dynamics. This means that  we can use the asymptotic behavior of L{\'e}vy distribution to study the properties of  far tails of the distribution of the position.
\item If $\mathcal{Z}\to \infty$ which is the short time limit, we have that the width of the L{\'e}vy distribution is narrow [see Eq.~\eqref{fadv101}] if compared with the width of Gaussian distribution. In this case, the   L{\'e}vy distribution approaches to a ``delta function''. As expected, we get the packet of spreading particles, following Gaussian distribution.
\end{itemize}
Notice that for a finite constant $\mathcal{Z}$    we  use the integral Eq.~\eqref{fadv101} to show the solution of fractional advection-diffusion equation (2) in the main text.

\begin{figure}[htb]
  \centering
  \includegraphics[width=13cm]{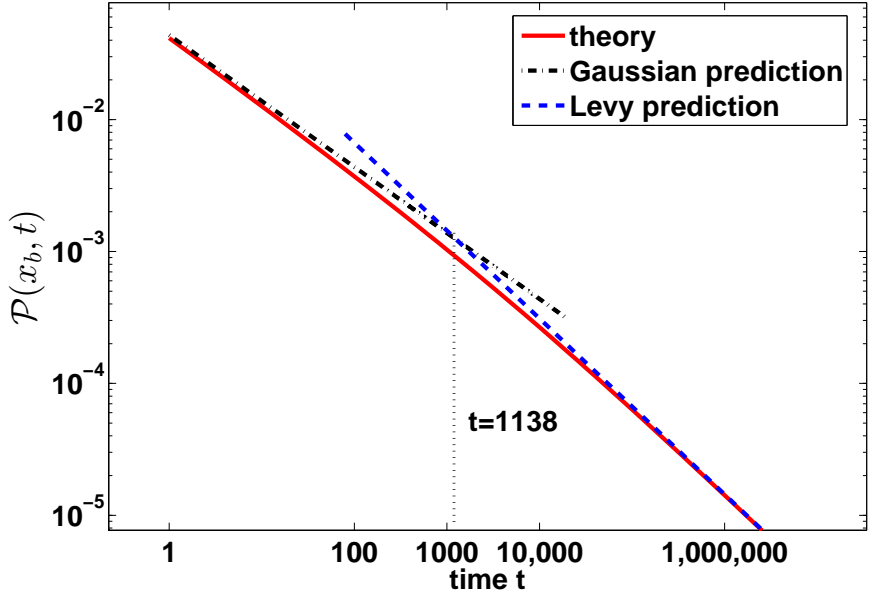}\\
  \caption{ Concentration $P(x_b,t)$ at changing site $x_b=Vt$ versus $t$. Solution of the fractional advection-diffusion equation, i.e., Eq.~\eqref{fadv101}, is the solid line. The dash-dotted  and the dashed lines are obtained from the corresponding Gaussian prediction valid for short times ($\mathcal{P}(x_b, t)\sim 1/\sqrt{4\pi Dt}$)  and L{\'e}vy  prediction ($\mathcal{P}(x_b, t)\sim L_{\beta}(0)/(St)^{1/\beta}$). Here we choose $\beta=1.5$, $D=41.7$, $V=3.3$, and $S=2.27$.}
\label{FADEQsmall}
\end{figure}

To demonstrate these properties, 
in Fig.~\ref{FADEQsmall} we plot the solution $\mathcal{P}(x_b,t)$ at site $x_b=Vt$ versus $t$, namely we focus on the probability of reaching the mean position. In other words, here $x_b$ is changing with time. 
Based on our setting in Fig.~\ref{FADEQsmall}, we choose $D= 41.7$, $V= 3.3$, and $S=2.27$ which are  transport constants of the fractional equation. If $t=10$, we have $\mathcal{Z}=26$. Thus we have $\mathcal{P}(x=V,t)\sim 1/\sqrt{4\pi Dt}$ for small $t$ since the packet of particles follows nearly Gaussian distribution. While, with the increase of the time $t$, as mentioned the Gaussian distribution fails, deviating from the solution of fractional advection-diffusion-asymmetry equation. Then at large times  the solution using the L{\'e}vy stable law is valid, i.e., $\mathcal{P}(x_b, t)\sim L_{\beta}(0)/(St)^{1/\beta}$.
This transition  is presented in Fig. \ref{FADEQsmall}.


\subsection{General fractional advection-diffusion-asymmetry equation}

All along the main text, for example Eq.~(5), we focus on the case when jump sizes have a finite  non-zero mean and variance.
We now further consider a general case, i.e., the displacement has a finite non-zero mean but infinite variance to extend the current Eq.~(2) in the main text. In Fourier space, the displacement follows
\begin{equation}\label{sd101}
\widetilde{f}(k)=\exp(-ika+(-ik)^\gamma)
\end{equation}
with $1<\gamma<2$,
from which we get the probability of reaching $X$ after exactly $N$ steps
\begin{equation}\label{levydisplacment}
P(X|N)=\frac{1}{N^{1/\gamma}}L_\gamma\left(\frac{X-aN}{N^{1/\gamma}}   \right).
\end{equation}
If $N=1$, Eq.~\eqref{levydisplacment} reduces to $f(x)$ in real space.
Utilizing Eqs.~(4) and \eqref{levydisplacment}, we obtain
\begin{equation}
\mathcal{P}_{{\rm CTRW}}(X,t)\sim \frac{1}{(t/\bar{t})^{1/\beta}}\int_0^\infty L_\beta\left(\frac{N-t/\langle\tau\rangle}{(t/\bar{t})^{1/\beta}}\right)\frac{1}{N^{1/\gamma}}L_\gamma\left(\frac{X-aN}{N^{1/\gamma}}\right){\rm d}N.
\end{equation}
Note that here $N$ is treated as a continuous variable. 
Changing variables of the above integral yields
\begin{equation}
\mathcal{P}_{{\rm CTRW}}(X,t)\sim \int_{-\infty}^\infty L_\beta(y)\frac{1}{(t/\langle\tau\rangle+y(t/\bar{t})^{1/\beta})^{1/\gamma}}L_\gamma\left(\frac{X-at/\langle\tau\rangle-ay(t/\bar{t})^{1/\beta}}{(t/\langle\tau\rangle+y(t/\bar{t})^{1/\beta})^{1/\gamma}}\right){\rm d}y
\end{equation}
and performing Fourier transform with respect to $X$ leads to
\begin{equation}\label{shdh102}
\widetilde{\mathcal{P}}_{{\rm CTRW}}(k,t)\sim \exp\left(-ika\frac{t}{\langle\tau\rangle}+(-ik)^\gamma \frac{t}{\langle\tau\rangle}\right)\int_{-\infty}^\infty L_\beta(y)\exp\left(-ikay\left(\frac{t}{\bar{t}}\right)^{1/\beta}+(-ik)^\gamma y\left(\frac{t}{\bar{t}}\right)^{1/\beta}\right){\rm d}y.
\end{equation}
We are interested in the statistics of $X$ in the long time limit.
For small $k$, Eq.~\eqref{shdh102} reduces to 
\begin{equation}\label{eqst1012}
\widetilde{\mathcal{P}}_{CTRW}(k,t)\sim \exp\left(-ika\frac{t}{\langle\tau\rangle}+(-ik)^\gamma \frac{t}{\langle\tau\rangle}+(-ik)^\beta\frac{a^\beta t}{\bar{t}}\right).
\end{equation}
Thus, the corresponding general fractional advection-diffusion equation reads 
\begin{equation}\label{eqst1013}
\frac{\partial }{\partial t }\mathcal{P}=-\frac{a}{\langle\tau\rangle}\frac{\partial }{\partial x }\mathcal{P}+\frac{1}{\langle\tau\rangle}\frac{\partial^\gamma}{\partial(-x)^\gamma}\mathcal{P}+\frac{a^\beta}{\bar{t}}\frac{\partial^\beta}{\partial(-x)^\beta}\mathcal{P}.
\end{equation}
In particular, when $\gamma=2$ in Eq.~\eqref{eqst1013}, we get Eq.~(2) in the main text which describes the case of Gaussian displacements. It can be seen that the method given in the main text is valid for a vast number of models. As mentioned before, we have two ways to derive Eq.~\eqref{eqst1013}. The first approach is from the well-known L{\'e}vy flight, i.e., the asymptotic displacement captures a non-zero mean and an infinite variance with two heavy tails. The second way dealing  with Eq.~\eqref{eqst1013} is a CTRW framework, where the waiting time has a fat tail with a finite mean and an infinite variance, and the distribution of the displacement has only one fat tail. It can be seen that the positional distribution for both approaches is the same.
The interesting point is that the types of particles' trajectories behind Eq.~\eqref{eqst1013} are totally different, see the discussion in the main text.

\section{Fractional advection-diffusion-asymmetry equation in two dimensions}\label{18ctrwSecttwoD10}
Now we study the advection-diffusion  equation in two dimensions and present  a  generalization of Eq. (2) in the main text. We further use a CTRW formalism in  two dimensions to explain the meaning of the equation.

Motivated by previous studies of the CTRW,  we consider here the probability density function capturing a fat tail
\begin{equation}\label{ldeq3201111}
\begin{array}{ll}
\psi(\tau)=\left\{
          \begin{split}
            &0, & \hbox{$\tau<\tau_0$;} \\
            &\beta\frac{(\tau_0)^\beta}{\tau^{1+\beta}}, & \hbox{$\tau\geq\tau_0$}
          \end{split}
        \right.
        \end{array}
\end{equation}
with $1<\beta<2$. Clearly, the waiting time $\tau$ has an finite mean $\langle\tau\rangle$ but a infinite variance. In our simulations, we choose $\tau_0=0.1$ and $\beta=3/2$ and hence $\langle\tau\rangle=0.3$. Thus, if the observation time $t=1000$, the average number of renewals is $\langle N\rangle\sim t/\langle \tau\rangle \simeq 3333$. 

The joint  PDF of jump length follows 
\begin{equation}\label{18ctrwSecttwoD102}
\begin{split}
  f(x,y) &=\frac{1}{\sqrt{2(\sigma_x)^2\pi}}\exp\left(-\frac{(x-a_x)^2}{2(\sigma_x)^2}\right)\times \frac{1}{\sqrt{2(\sigma_y)^2\pi}}\exp\left(-\frac{y^2}{2(\sigma_y)^2}\right),
\end{split}
\end{equation}
where $a_x,~\sigma_x$, $\sigma_y\neq 0$ are constants.  This indicates that the drift is only in $x$ direction. 
In double Fourier spaces, $x\to k_x$ and $y\to k_y$, we get
\begin{equation}\label{18ctrwSecttwoD103}
  \widetilde{f}(k_x,k_y)=\exp\left(-ik_xa_x-\frac{1}{2}(\sigma_x)^2(k_x)^2-\frac{1}{2}(\sigma_y)^2(k_y)^2\right),
\end{equation}
which gives the PDF of finding the particle on site $(X,Y)$ after exactly $N$ steps by taking the inverse Fourier transform of $\widetilde{f}^N(k_x,k_y)$. 
Restarting from Eq.~(7) in the main text, in the long time limit, $\mathcal{P}(X,Y,t)_{{\rm CTRW}}$ becomes
\begin{equation}\label{18ctrwSecttwoD104}
 \mathcal{P}(X,Y,t)_{{\rm CTRW}}\sim\int_{-\infty}^\infty L_{\beta}(z)\frac{\exp\left(-\frac{(X-\frac{a_xt}{\langle\tau\rangle}-a_xz(\frac{t}{\overline{t}})^{1/\beta})^2}{2(\sigma_x)^2\frac{t}{\langle\tau\rangle}}\right)}{\sqrt{2\pi(\sigma_x)^2\frac{t}{\langle\tau\rangle}}}\frac{\exp\left(-\frac{Y^2}{2(\sigma_y)^2\frac{t}{\langle\tau\rangle}}\right)}{\sqrt{2\pi(\sigma_y)^2\frac{t}{\langle\tau\rangle}}}{\rm d}z.
\end{equation}
Taking double Fourier transforms with respect to $X$ and $Y$, respectively, we get a useful expression
\begin{equation}\label{18ctrwSecttwoD105}
 \widetilde{P}(k_x,k_y,t)_{{\rm CTRW}}\sim\exp\left(-(k_y)^2(\sigma_y)^2\frac{t}{2\langle\tau\rangle}-(k_x)^2(\sigma_x)^2\frac{t}{2\langle\tau\rangle}-ia_xk_x\frac{t}{\langle\tau\rangle}+\left(-ia_xk_x\left(\frac{t}{\overline{t}}\right)^{1/\beta}\right)^\beta\right).
\end{equation}
Note that if $k_x=k_y=0$, we can check that $\mathcal{P}(X,Y,t)_{{\rm CTRW}}$ is normalized for any time $t$. Similar to the calculation in the main text, we take the time derivative of this solution, perform the inverse Fourier transform and then find using Eqs.~\eqref{weyl3} 
and \eqref{18ctrwSecttwoD105}
\begin{equation}\label{18ctrwSecttwoD106}
\frac{\partial}{\partial t}\mathcal{P}(x,y,t)=\frac{(\sigma_y)^2}{2\langle\tau\rangle}\frac{\partial^2}{\partial y^2}\mathcal{P}(x,y,t)+\frac{(\sigma_x)^2}{2\langle\tau\rangle}\frac{\partial^2}{\partial x^2}\mathcal{P}(x,y,t)-\frac{a_x}{\langle\tau\rangle}\frac{\partial}{\partial x}\mathcal{P}(x,y,t)+\frac{(a_x)^\beta}{\overline{t}}~\frac{\partial^\beta}{\partial (-x)^\beta} \mathcal{P}(x,y,t).
\end{equation}

This is the fractional advection-diffusion  equation in two dimensions, where the symmetry breaking takes place only in the $x$ direction, where the bias is pointing to.
The solution of the above equation is plotted in Fig.~1 in the main text showing the packet of spreading particles. In this figure, we denote $a$ as $a_x$ to simplify our expression.
Note that Eq.~\eqref{18ctrwSecttwoD106} is not just valid for Gaussian displacement given in Eq.~\eqref{18ctrwSecttwoD104} but the displacement should  have a finite mean  and a finite variance. In particular, when $a_x=0$, the above equation reduces to the classical diffusion equation.
Clearly, the marginal density $\mathcal{P}(x,t)$ is  the corresponding one dimensional solution Eq.~(7) in the main text.

\section{Breakthrough curves}
Here the aim is to use the fractional advection-diffusion-asymmetry equation found in the main text to predict  breakthrough curves. For that, the first step is to obtain $\mathcal{P}(x,t)$ from Eq. (2) with time-dependent bias and then use it to compare with  simulations of CTRW breakthrough curves.

\subsection{Theory of propagator with time-dependent but piece
wise constant bias}
Motivated by \cite{Nissan}, we consider time-dependent bias determined by four stages.
We suppose that the rapid injection of particles is done immediately after starting observing the process. In other words,  the  initial condition of the particle is $\mathcal{P}(x,t=0)=\delta(x)$. As mentioned in the main text we simulate the spreading of the particles consisting of four states: (i) after the injection of the particles, they are moving with a constant bias which is determined by $a_1=a$, (ii) in the time interval $t_1<t<t_2$, we increase the bias sharply to $a_2=4a \gamma/(\gamma+1)$ with $\gamma\geq1/3$, (iii)  decrease the bias abruptly to $a_3=a/(2/3+\gamma)$, and (iv) then finally starting at $t_3$  return to the state (i) with $a_4=a$. Here $\gamma$ is a  constant that controls the  strength of bias or the average of ``velocity''. In particular, when $\gamma=1/3$, and hence all the states mentioned above are the same.  In Fourier space, the initial condition satisfies $\widetilde{\mathcal{P}}(k,0)=1$. 
In view of the special expression of $\widetilde{\mathcal{P}}(k,t)$,  $\widetilde{\mathcal{P}}(k,t)$ for different states can be cast as

\begin{equation}\label{aaeqfk301}
\widetilde{\mathcal{P}}(k,t)=\left\{
  \begin{array}{ll}
    \exp(-c_{11}k^2-ic_{12}k+c_{13}(-ik)^\beta), & \hbox{$0<t\leq t_1$;} \\
   \exp(-c_{21}k^2-ic_{22}k+c_{22}(-ik)^\beta), & \hbox{$t_1<t\leq t_2$;} \\
   \exp(-c_{31}k^2-ic_{32}k+c_{33}(-ik)^\beta)  , & \hbox{$t_2<t\leq t_3$;} \\
   \exp(-c_{41}k^2-ic_{42}k+c_{43}(-ik)^\beta) , & \hbox{$t_3<t$}
  \end{array}
\right.
\end{equation}
with
\begin{equation}\label{aaeqfk302}
c_{m1}=\left\{
  \begin{array}{ll}
   t\frac{\sigma^2}{2\langle\tau\rangle} , & \hbox{$m=1$;} \\
   t\frac{\sigma^2}{2\langle\tau\rangle}, & \hbox{$m=2$;} \\
   t\frac{\sigma^2}{2\langle\tau\rangle}, & \hbox{$m=3$;} \\
   t\frac{\sigma^2}{2\langle\tau\rangle} , & \hbox{$m=4$,}
  \end{array}
\right.
\end{equation}

\begin{equation}\label{aaeqfk303}
c_{m2}=\left\{
  \begin{array}{ll}
   a_1\frac{t}{\langle\tau\rangle}, & \hbox{$m=1$;} \\
   \left[a_1t_1+a_2(t-t_1)\right]\frac{1}{\langle\tau\rangle}, & \hbox{$m=2$;} \\
  \left[a_1t_1+a_2(t_2-t_1)+a_3(t-t_2)\right]\frac{1}{\langle\tau\rangle}, & \hbox{$m=3$;} \\
  \left[a_1t_1+a_2(t_2-t_1)+a_3(t_3-t_2)+a_4(t-t_3)\right]\frac{1}{\langle\tau\rangle} , & \hbox{$m=4$,}
  \end{array}
\right.
\end{equation}
and
\begin{equation}\label{aaeqfk304}
c_{m3}=\left\{
  \begin{array}{ll}
   a_1^\beta t\frac{1}{\overline{t}}, & \hbox{$m=1$;} \\
   \left[a_1^\beta t_1+a_2^\beta(t-t_1)\right]\frac{1}{\overline{t}}, & \hbox{$m=2$;} \\
  \left[a_1^\beta t+a_2^\beta(t_2-t_1)+a_3^\beta(t-t_2)\right] \frac{1}{\overline{t}}, & \hbox{$m=3$;} \\
 \left[a_1^\beta t_1+a_2^\beta(t_2-t_1)+a_3^\beta(t_3-t_2)+a_4^\beta(t-t_3)\right] \frac{1}{\overline{t}}, & \hbox{$m=4$.}
  \end{array}
\right.
\end{equation}
Here recall that $\bar{t} =\langle\tau\rangle^{1 +\beta}/[(\tau_0)^\beta|\Gamma(1 - \beta)|]$ which is fixed, since in our simulations of the CTRW we only change  the bias. 
The main idea of the analytical  calculation is that the final position of each stage will be treated as an ``initial condition'' for the next stage.
In particular, if $a_1=a_{2}=a_{3}=a_{4}$, Eq.~\eqref{aaeqfk301} reduces to a constant bias case calculated in the main text. 
For the four-stage process, from Eq.~\eqref{aaeqfk301} the solution is of the form
\begin{equation}\label{aaeqfk305}
\mathcal{P}(x,t)=\int_{-\infty}^{\infty}\frac{1}{\sqrt{4\pi c_{m1}}}\exp\left[-\frac{(x-y-c_{m2})^2}{4c_{m1}}\right]\frac{1}{(c_{m3})^{1/\beta}}L_{\beta}\left[\frac{y}{(c_{m3})^{1/\beta}}\right]{\rm d}y
\end{equation}
with $m=1,2,3,4$ being the number of the state. This is plotted in  Fig.~ \ref{FixTxi10si5} for different times $t$. Note that when $0<t<t_1$, the solution Eq.~\eqref{aaeqfk305} reduces to  Eq.~(7) in the main text.

\subsection{Breakthrough curves}
 As mentioned in the main text breakthrough curves are measured considering  a source that passes through the absorbent fixed bed sample, which is a method to analyze
the adsorption properties of tracers in  porous materials. An example of experiments is the transport through layers of different
media, see Refs.
\cite{Berkowitz2009shcer,Nissan}.
With Eq.~\eqref{aaeqfk305} we constructed the analytical solution presented in Fig. 2 of the main text. In Fig. 2 we  use $\gamma=10$ for curve A, and further choose $a=1$, so there we have in the four stages $a=\{1,3.6,0.09,1\}$. Curve B in Fig. 2 corresponds to the case $\gamma=1/3$, where we get $a=\{1,1,1,1\}$. For the time interval of each state we use $t_1=t_2-t_1=t_3-t_2=100$ and $t-t_3=700$. Note that the detection wall is set on $x_b=1800$.
In addition to Fig.~2 in the main text here we present results  for the density with $t_1=t_2-t_1=t_3-t_2=10$ as this shows the quick convergence to our theory; see Fig. \ref{FixXshortTime}. Here the  time interval  is $10$ for the first three stages, given that $\langle\tau\rangle=0.3$ we have roughly $33$ steps in  each  of the first three time intervals. The figure illustrates that even for these relatively short time intervals the approximation works nicely. 

\begin{figure*}[htp]
    \begin{minipage}[h]{1\linewidth}
\begin{center}
 \includegraphics[width=1\textwidth]{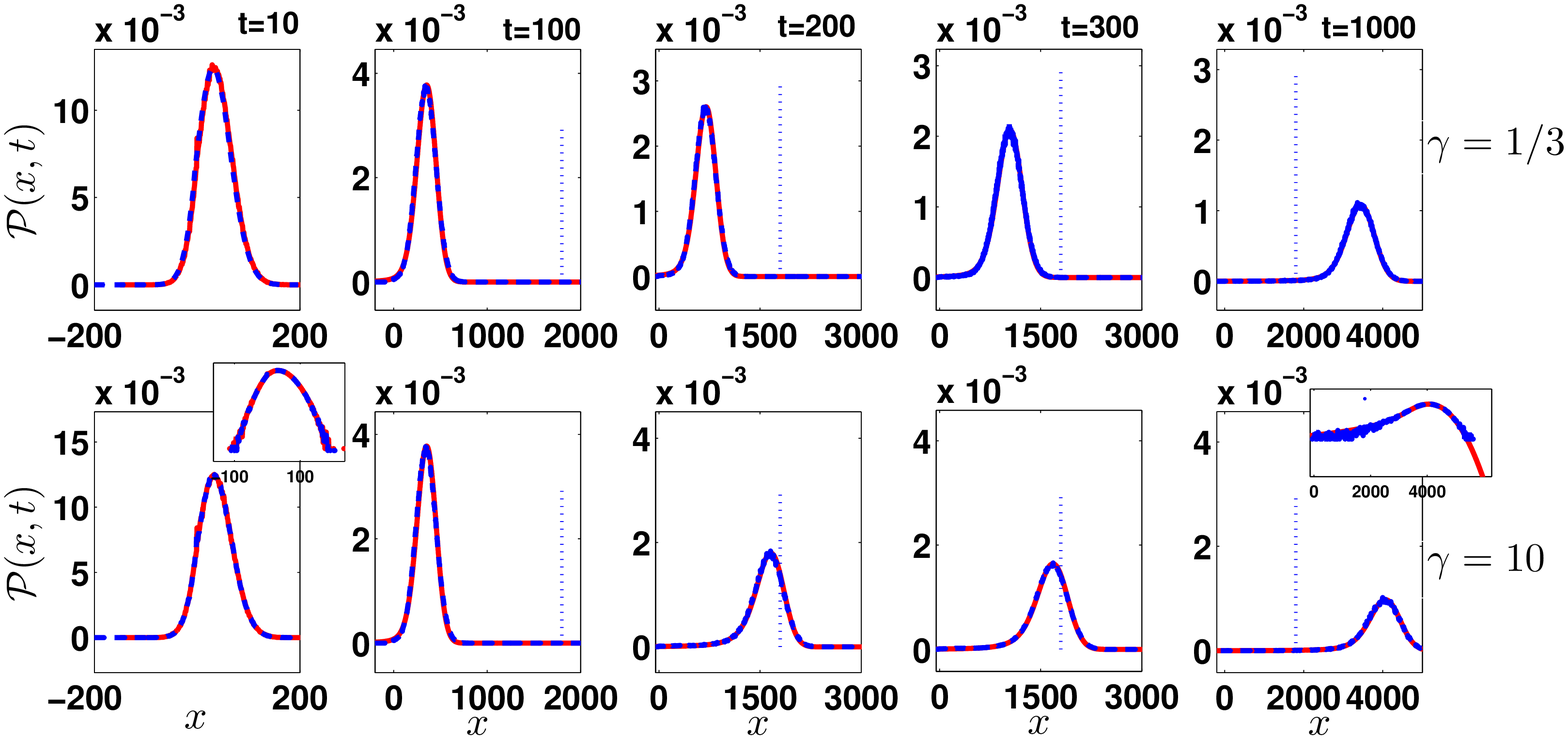}
\end{center}
    \caption{Simulations of propagators for different states with theoretical predictions Eq.~\eqref{aaeqfk305}. Waiting times of particles are drawn from Eq.~\eqref{ldeq3201111} and displacements generated according to $f(x)=\exp(-(x-a)^2/2\sigma^2)/\sqrt{2\pi\sigma^2}$, where $a$ is time dependent obtained for $\gamma=1/3~({\rm the~first~ row}),~10$~(the~second).
    The red solid lines are the theory and the blue dashed lines are the simulation results generated from $2\times10^6$ trajectories. The time interval of the first three states is $100$ and the total time of the last state is $700$. Note that here  it is not easy to see the asymmetry of the packet of spreading particles in a linear-linear plot for large $t$, at least this figure can be treated as an optical illusion. However, semi-log scales are presented to show the symmetric properties of the packet and the heavy tail; see the left and the right insets.    The parameters are $\beta=3/2$, $\tau_0=0.1$,  and~  $\sigma=5$, which are the same as Fig.~2 in the main text. 
 }
    \label{FixTxi10si5}
    \end{minipage}
\end{figure*}

\begin{figure}[htb]
  \centering
  \includegraphics[width=13cm]{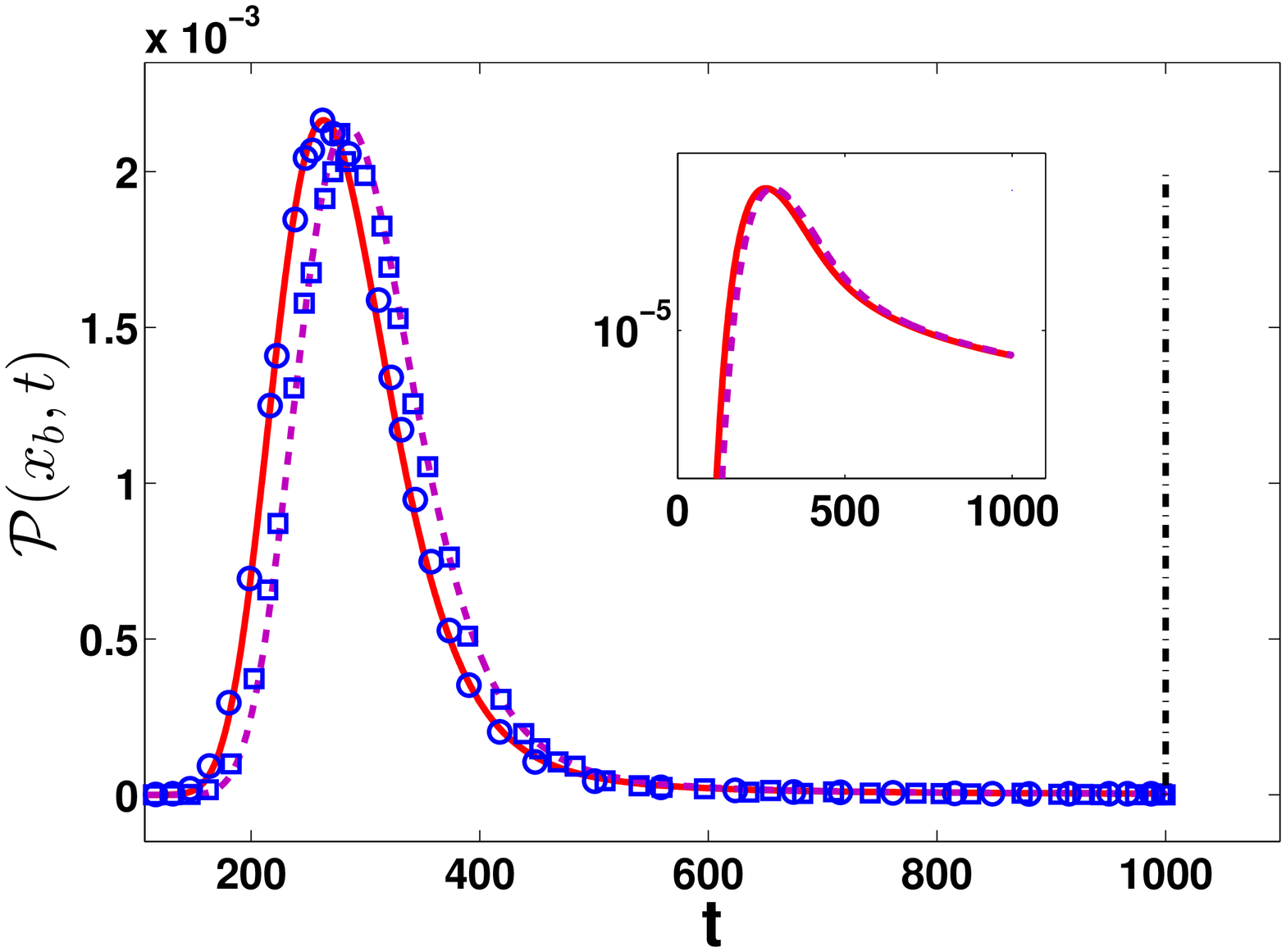}\\
  \caption{A plot of breakthrough curves for a weak bias.
Here we choose $\beta=3/2$, $\tau_0=0.1$, $t_1=10$, $t_2=20$, $t_3=30$, $t=1000$, $\gamma=1/3$ (solid lines), and $\gamma=10$ (dashed lines). The detection wall for breakthrough curves is $x_b=10^3$; see the dash-dotted line.  For both cases,
after reaching a peak, we see a slow decay of the far right tail to zero,
if compared to the left one (see the inset). Clearly
the theoretical predictions Eq.~\eqref{aaeqfk301} (the solid and the dashed lines) are consistent with simulations results plotted by the symbols.  }
\label{FixXshortTime}
\end{figure}

%

\begin{figure}[htb]
  \centering
  \includegraphics[width=18cm]{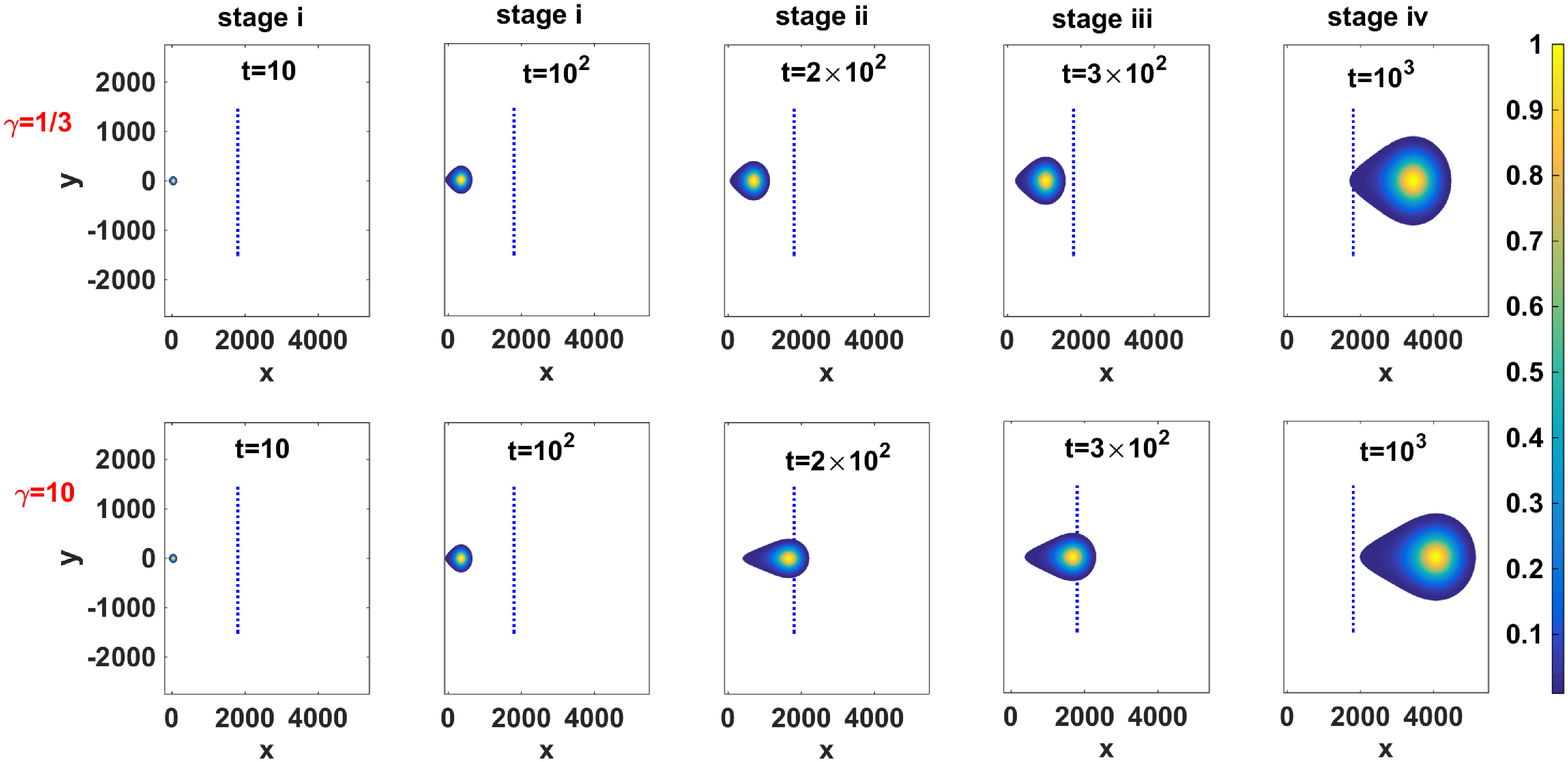}\\
  \caption{Theoretical prediction $\mathcal{P}(x,y,t)/\mathcal{P}_{\max}$ showing   spatial contour maps under the  time-dependent bias in two dimensions, where $\mathcal{P}_{\max}=\max_{x,y\in (-\infty,\infty)\times(-\infty,\infty)}(\mathcal{P}(x,y,t))$.
  The colorbar represents the relative concentration, $\mathcal{P}(x,y,t)/{\mathcal{P}_{\max}}$. Here fours stages are $0<t<100$, $100<t<200$, $200<t<300$, and $300<t<1000$.
  The joint PDF of displacements follows Eq.~\eqref{18ctrwSecttwoD102} with a changing $a_x$, namely we have $a_x=\{1, 3.6, 0.09, 1\}$~~($a_x=\{1, 1, 1, 1\}$) when $\gamma=10$~ ($\gamma=1/3$); see Eq.~\eqref{sdje101}. The theoretical results are obtained from Eqs.~\eqref{aaeqfk301} and \eqref{18ctrwSecttwoD105}, namely
  $\mathcal{P}(x,y,t)$  factorizes into a product of  Eq.~\eqref{aaeqfk305} and $\exp(-y^2/2(\sigma_y)^2(t/\langle\tau\rangle))/\sqrt{2\pi(\sigma_y)^2(t/\langle\tau\rangle)}$. In our setting, we choose $\sigma_x=\sigma_y=5$, and
the other parameters are the same as Fig.~\ref{FixTxi10si5}. 
}
\label{2dFADeq}
\end{figure}

Similarly, one can deal with  time-dependent bias in two dimensions using Eq.~\eqref{18ctrwSecttwoD105} like Eq.~\eqref{aaeqfk301}. In Fig.~\ref{2dFADeq}, 
the time-dependent bias in two dimensions is investigated and the asymmetry packet with respect to $x$ is clearly observed. Here we have the drift in the $x$ direction and no bias in the $y$ direction, i.e.,
\begin{equation}\label{sdje101}
a_x=\left\{
\begin{array}{ll}
a, & \hbox{$0<t\leq t_1$;} \\
\frac{4a}{\gamma+1}, & \hbox{$t_1<t\leq t_2$;} \\
\frac{a}{\gamma+\frac{2}{3}}, & \hbox{$t_2<t\leq t_3$;}\\
a, & \hbox{$t_3<t$.}
\end{array}
\right.
\end{equation}
It can be seen that Figs.~\ref{FixTxi10si5} and \ref{2dFADeq} are  complementary and yield a better understanding of breakthrough curves and the asymmetric packets of spreading particles.

\end{appendices}

\end{document}